\documentclass[preprint2]{aastex}
\usepackage{graphics}
\usepackage{epsfig}
\usepackage {url} 
\usepackage{gensymb}
\def\kms{\mbox{~km~s$^{-1}$}}
\def\kpc{\mbox{~kpc}}

\def\kpch{\mbox{$~h^{-1}$ kpc}}

\def\LCDM{\mbox{$\Lambda$CDM}}
\def\Mpc{\mbox{~Mpc}}

\def\Mpch{\mbox{$~h^{-1}$ Mpc}}
\def\Msun{\mbox{~M$_\odot$}}
\def\Msunh{\mbox{$~h^{-1}$M$_\odot$}}

\def\Rvir{R_{\rm vir}}
\def\M2vir{M_{\rm 2vir}}
\def\Mvir{M_{\rm vir}}
\def\Mtot{M_{\rm tot}}
\def\MDM{M_{\rm DM}}
\def\Mbar{M_{\rm bar}}
\def\fbar{f_{\rm bar}}

\def\Omegabar{\Omega_{\rm bar}}
\def\Omegamatter{\Omega_{\rm matter}}
\def\vc{V_{\rm circ}}
\def\vcirc{V_{\rm circ}}
\def\vmax{V_{\rm max}}
\def\vacc{V_{\rm acc}}
\def\vten{V_{10}}
\def\rs{r_{\rm s}}

\begin{document}
\shortauthors{TRUJILLO-GOMEZ ET AL.}  
\shorttitle{} 
\title{Galaxies in \boldmath$\LCDM$ with Halo Abundance Matching: 
luminosity-velocity relation, baryonic mass-velocity relation,
velocity function and clustering}

\author{Sebastian Trujillo-Gomez$^1$, Anatoly Klypin$^1$, Joel
  Primack$^2$, and Aaron J. Romanowsky$^3$}

\affil{$^1$Astronomy Department, New Mexico State University,
MSC 4500, P.O.Box 30001, Las Cruces, NM 88003-8001 USA}

\affil{$^2$Department of Physics, University of California at Santa
  Cruz, Santa Cruz, CA 95064 USA}

\affil{$^3$UCO/Lick Observatory, University of California at Santa Cruz, Santa Cruz, CA 95064
USA}



\begin{abstract}

  It has long been regarded as difficult if not impossible for a
  cosmological model to account simultaneously for the galaxy
  luminosity, mass, and velocity distributions.  We revisit this issue
  using a modern compilation of observational data along with the 
  best available large-scale cosmological simulation of dark matter. We find
  that the standard cosmological model, used in conjunction with halo
  abundance matching (HAM) and simple dynamical corrections, fits -- at least on average --
  all basic statistics of galaxies with circular velocities
  $\vcirc > 80\kms$ calculated at a radius of $\sim$10~kpc. Our primary observational
  constraint is the luminosity-velocity relation -- which generalizes
  the Tully-Fisher and Faber-Jackson relations in allowing all types
  of galaxies to be included, and provides a fundamental benchmark to
  be reproduced by any theory of galaxy formation. We have compiled
  data for a variety of galaxies ranging from dwarf
  irregulars to giant ellipticals.   The data present a clear monotonic
  luminosity-velocity relation from $\sim$50\kms\ to $\sim$500\kms,
  with a bend below $\sim$~80\kms\ and a systematic offset between
  late- and early-type galaxies.  For comparison to theory, we employ our new $\LCDM$ ``Bolshoi'' simulation of dark matter,
  which has unprecedented mass and force resolution over a large
  cosmological volume, while using an up-to-date set of cosmological
  parameters.  We use halo abundance matching to assign
  rank-ordered galaxy luminosities to the dark matter halos, a
  procedure that automatically fits the empirical luminosity function
  and provides a predicted luminosity-velocity relation that can be
  checked against observations.  The adiabatic contraction of dark
  matter halos in response to the infall of the baryons is included as
  an optional model ingredient.  The resulting predictions for the
  luminosity-velocity relation are in excellent agreement with the
  available data on both early-type and late-type galaxies for the
 luminosity range from $M_r=-14$ to $M_r=-22$.  We also compare our predictions for
  the ``cold'' baryon mass (i.e., stars and cold gas) of galaxies as a function of circular velocity
  with the available observations, again finding a very good agreement.  The predicted
  circular velocity function is also in agreement with the galaxy velocity
  function from 80 to 400 km s$^{-1}$, using the HIPASS survey for
  late-type and SDSS for early-type galaxies.  However, in accord
  with other recent results, we find that the dark matter halos with
  $V_{\rm circ}<80 \kms$ are much more abundant than observed
  galaxies with the same $V_{\rm circ}$. Finally, we find that the two-point correlation function of bright galaxies in our model matches very well the results from the final data release of the SDSS, especially when a small amount of scatter is included in the HAM prescription.

\end{abstract}



\section{Introduction}
\label{sec:intro}

The cosmological constant + cold dark matter ($\LCDM$) model is
the reigning paradigm of structure formation in the universe. The
presence of large amounts of dark mass in the surroundings of galaxies
and within galaxy clusters has been established firmly using dynamical
mass estimates that include spiral galaxy rotation curves, velocity
dispersions of galaxies in clusters and x-ray emission measurements of
the hot gas in these systems, as well as strong and weak lensing of background
galaxies. The $\LCDM$ model also correctly predicts the details of the
temperature and polarization of the cosmic background radiation \citep{komatsu10}.
A few issues remain where the model and the observations are
either hard to reconcile or very difficult to compare \citep{primack09}. 
Examples of this are the so-called missing satellites problem
\citep{klypin99,moore99,Bullock00,willman04,Maccio10} and the cusp/core
nature of the central density profiles of dwarf galaxies
\citep{flores94,moore94,deBlok97,Valenzuela07,Governato10,deBlok10}.  

An outstanding challenge for the $\LCDM$ model that we address here 
is to reproduce the observed abundance of galaxies as a function of
their overall properties such as dynamical mass, luminosity, stellar
mass, and morphology, both nearby and at higher redshifts.  A successful
cosmological model should produce agreement with various
observed galaxy dynamical scaling laws such as the Faber-Jackson 
\citep{faber76} and Tully-Fisher \citep{tully77} relations.

Making theoretical predictions for properties of galaxies that can be
tested against observations is difficult.  While dissipationless
simulations can provide remarkably accurate predictions of various
properties of dark matter halos, they do not yet make secure
predictions about what we actually observe -- the distribution and
motions of stars and gas.  We need to find a common ground where
theoretical predictions can be confronted with observations.  In this
paper we use three statistics to compare theory and observations: (a)
the luminosity - circular velocity (LV) relation, (b) the baryonic
Tully-Fisher relation (BTF), and (c) the circular velocity function
(VF).

 In all three cases we need to estimate the circular velocity (a
 metric of dynamical mass) at some distance from the center of each
 dark matter halo that hosts a visible galaxy. Unfortunately, theory
 cannot yet make accurate predictions for the central regions of
 galaxies because of uncertain baryonic astrophysics.  As a
 compromise, we propose to use the distance of 10~kpc. Measurements of
 rotational or circular velocities of galaxies either exist for this
 distance or can be approximated by extrapolations. At the same time,
 theoretical predictions at 10~kpc are also simplified because they
 avoid the complications of the central regions of galaxies.

Our LV relation is a close cousin of the Tully-Fisher (TF) relation and,
indeed, we will use some observational results used to construct the traditional Tully-Fisher relation. However, there are substantial
differences between the TF and the LV relations. The standard
Tully-Fisher relation tells us how quickly spiral disks rotate for
given luminosity. The rotation velocity is typically measured at 
2.2 disk scale lengths \citep[e.g.,][]{courteau07}, where the ``cold''
baryons (i.e., stars and cold gas)
contribute a substantial fraction of the mass. Instead, at the 10~kpc
radius used here for the LV relation, the dark matter is the dominant
contribution to the mass in all but the largest galaxies. 
More importantly, the LV relation includes not 
only spiral galaxies, but all morphological types. 
Thus, {\it the LV relation is the relation between the galaxy luminosity and the total
mass inside the 10~kpc fiducial radius.}
 
In order to make theoretical predictions for the LV relation, we need
to estimate the luminosity of a galaxy expected to be hosted by a dark
matter halo (including those that are substructures of other halos). There are different ways to make those predictions.
Cosmological $N$-body+gasdynamics simulations will eventually be an
ideal tool for this.  However, simulations are still far from
achieving the resolution and physical understanding necessary to
correctly model the small scale physics of galaxy formation and
evolution.  Early simulations had problems reproducing the TF relation
\citep[e.g.,][]{navarro00}.  \citet{eke01} could reproduce the slope
of the TF relation, but created disks that were too faint by about
$0.5$ magnitudes in the $I$-band for any given circular
velocity. Recently the situation has improved. For example,
\citet{governato07} produced disk galaxies spanning a decade in mass
that seem to fit both the $I$-band TF relation and the baryonic TF
relation very well, as well as the observed abundance of Milky
Way-type satellites. Most recently, \citet{guedes11} have produced perhaps the best match yet to a Milky Way-type galaxy in $\LCDM$ using a high-resolution smooth particle hydrodynamics simulation with a high density threshold for star formation.

Making predictions for a large ensemble of simulated galaxies is yet
another challenge.  Semi-analytical models (SAMs) are a way to make
some progress in this direction.  These models have the advantage of
producing large-number galaxy statistics.  They typically include many
free parameters controlling the strength of the various processes that
affect the build-up of the stellar population of a galaxy (i.e.,
cooling, star formation, feedback, starbursts, AGNs, etc.).
Unfortunately, these normalizing parameters can be difficult to
constrain observationally \citep[e.g.,][]{somerville99,benson10}.  The
models aim to reproduce the observed number distributions of galaxies
as a function of observables such as luminosity, stellar mass, cold
gas mass, and half-light radius, along with scaling laws such as the
Tully-Fisher relation and the metallicity-luminosity relation.

Early SAMs suffered from serious defects.  The models of
\citet{kauffmann93} were normalized using the observed TF relation
zero-point, which resulted in a luminosity function with a very steep
faint-end. On the other hand, models such as those of \citet{cole94}
were normalized to reproduce the observed ``knee'' in the LF but this
resulted in a large offset in the TF relation zero-point. Later models
have shown moderate success in reproducing either the luminosity
function \citep{benson03} or the TF relation \citep{somerville99}, but
it has been difficult to match both simultaneously when rotation
curves are treated realistically \citep{cole00}.  \citet{benson03}
used a combination of disk and halo reheating to obtain reasonable
agreement with the observed LF except at the faint end, where they
still overpredict the number of dwarf galaxies. If the WMAP 5-year
cosmology \citep{komatsu09} were used, their models would also produce
too many very bright galaxies. The TF relation they obtain has the
correct slope but their disks are too massive at any given
luminosity. Most recently, \citet{benson10} used a sophisticated
version of their GALFORM semi-analytic model to obtain sets of
parameters that minimize the deviations from twenty one observational
datasets including the LFs in several bands and at different
redshifts, the TF relation, the average star formation rate as a
function of redshift, clustering and metallicities among many
others. Not surprisingly, even their best model has difficulty fitting
such a large number of simultaneous constraints. In particular, the LF
in the $K$ band overpredicts the number of dwarf galaxies by almost an
order of magnitude at the faint end, while the LFs at high redshift
consistently overpredict the abundance of all galaxies. In addition,
the halos they obtain contain too many satellite galaxies, resulting
in too strong a galaxy two-point correlation in the one-halo
regime. The Tully-Fisher relation of their best fit model also shows a
systematic offset of about $20-40\kms$ towards higher circular
velocities for any given luminosity when compared to observations.

Recent high-resolution $N$-body cosmological simulations such as 
\citet{MSI,klypin10} have volumes large
enough to obtain the mass function of dark matter (DM) halos, but there
is no direct way to compare it to observational measurements of the
luminosity or stellar mass functions of galaxies. A new technique
recently emerged that allows us to bridge the gap between dark
matter halos and galaxies. It is commonly referred to as abundance matching
\citep{Kravtsov04,tasitsiomi04,Vale04,conroy06,Conroy09,Guo10,behroozi10}.
Halo abundance matching (HAM)
resolves the issue of connecting observed galaxies to simulated dark
matter (DM) halos by setting a one-to-one correspondence between red-band
luminosity and dynamical mass: more luminous galaxies are assigned to
more massive halos.  By construction, it reproduces the observed
luminosity function. It also reproduces the scale dependence of galaxy
clustering over a range of epochs \citep{conroy06,Guo10}. When abundance
matching is used for the observed stellar mass function \citep{li09},
it gives a reasonably good fit to the lensing results
\citep{Mandelbaum06} on the relation between the stellar mass and the
virial mass \citep{Guo10}. \citet{Guo10} also tried to reproduce the
observed relation between the stellar mass and the circular velocity
with partial success: there were deviations in both the shape and the
amplitude. At circular velocities $V_c=100-150 \kms$ the predicted
circular velocity was $\sim 25$\% lower than the observed one. They
argued that the disagreement is likely due to the fact that they did
not include the effect of cold baryons. Below we show that this is indeed the
case.

The paper is structured in the following
way. Section~\ref{sec:observations} describes in detail the
observational samples used to compare with the results of our
analysis. Section~\ref{sec:sim} briefly describes our new Bolshoi
simulation \citep{klypin10} and compares it to other large
cosmological simulations. In section~\ref{sec:halo} we describe some
characteristics of dark matter halos.  Section~\ref{sec:procedure}
describes the abundance matching method used to relate observed
galaxies to the DM halos in the Bolshoi simulation and explains the
procedure used to measure key quantities such as the circular velocity
for these model galaxies. Section~\ref{sec:observations} describes in
detail the observational samples used to compare with the results of
our analysis.  Section~\ref{sec:results} shows the LV relation, the
baryonic Tully-Fisher relation, the galaxy circular velocity
function and the galaxy two-point correlation function obtained using
our model and compares them to the observations described in
Section~\ref{sec:observations}. A brief comparison with related
results in the literature is given in
Section~\ref{sec:compare}. Section~\ref{sec:discussion} presents a
discussion of our results and Section~\ref{sec:conclusions} summarizes
them.

\section{Observational Data}
\label{sec:observations}

\subsection{Late-type galaxies}
\label{sec:LTG}

We use several datasets to construct the LV relation
for observed galaxies.  \citet{springob07} compiled a template $I$-band
Tully-Fisher sample of 807 spiral galaxies of types Sa-Sd in order to
calibrate distances to $\sim 4000$ galaxies in the local universe. Template galaxies were chosen to be members of nearby clusters in
order to minimize distance errors. Their photometry contains distance
uncertainties so the scatter should be taken cautiously and only as an
upper limit to the intrinsic TF scatter. Circular velocities were
obtained using HI line synthesis observations or optical H$\alpha$
rotation curves when those were not available. The maximum circular
velocity was obtained by using a model fit to the observed
profiles. Since the authors
correct for the effects of turbulence by subtracting $6.5\kms$
linearly from the velocity widths, it was necessary to de-correct them
by adding this term back in to obtain the true circular velocities.

The \citet{pizagno07} sample was selected from the Sloan Digital Sky
Survey (SDSS) \citep{york00}. It is one of the most complete and
unbiased samples available of H$\alpha$ rotation curves of disk
galaxies and was studied in an attempt to
accurately measure the intrinsic scatter in the TF
relation. Luminosities were taken from SDSS $r$-band photometry,
yielding the best  match with the luminosities assigned to our
model galaxies. For this sample we used the asymptotic value of the
rotation velocity they obtained using a functional fit to the rotation
curves.

In order to test the predictions of the $\LCDM$ model using abundance
matching ($\LCDM$ + HAM for short)  with the largest dynamic
range possible, we included in our comparison the latest Tully-Fisher
dwarf galaxy sample studied by \citet{geha06}. Their sample consists
of about 110 late-type galaxies with luminosities measured in the
$r$-band and rotation velocities measured using HI emission. 

The three samples above constitute our major observational dataset. We
further cut them by selecting galaxies with high inclinations ($i>45 \degree$ or axis ratio $b/a > 0.7$) to minimize uncertainties due to
projection effects. Additionally, we include only galaxies with better
than $10 \%$ accuracy in the measurement of the maximum circular
velocity. These cuts leave a total of 972 galaxies in the major sample.

For comparison with the datasets mentioned above, we include other smaller samples found in the
literature. The sample of \citet{blanton08} was also obtained from SDSS and is comprised of only isolated galaxies
with high inclinations. The HI galaxy sample used by \citet{sakai00}
was selected to have small scatter for use as a distance
calibrator. It is important to note that while the fit shown here
minimizes both the errors in rotation velocity and in luminosity, it
may be artificially shallow due to selection
effects.

Certain assumptions about galaxy colors had to be made in order to
convert the different observational samples to the $^{0.1}r$-band
measurements we chose for our model. In order to convert the $I$-band
luminosities measured by \citet{springob07} to the $r$-band, we
cross-referenced their data with the sample of \citet{pizagno07} and
used the median $(r-I)$ colors of the galaxies present in both
catalogs. To convert from the $R$-band magnitudes of \citet{sakai00}
to the SDSS $^{0.1}r$-band we used the transformation equations
obtained by \citet{lupton05} along with the typical $^{0.1}(r-i)$
color of disk galaxies in the SDSS sample studied by
\citet{blanton03a}. In addition, for redshift zero data sets, the
k-correction given in \citet{blanton03b} was used to convert from
$z=0$ to $z=0.1$ photometric bands.

Lastly, since the obscuring effect of dust extinction as a function of
disk inclination is corrected for in Tully-Fisher samples but not in
observed LF estimates, we had to de-correct the luminosities of the
spiral galaxies in all of the TF samples. To do so, we estimated and added the median
extinction in the $r$-band as a function of rotation velocity using
the method and sample employed by \citet{pizagno07}. This correction is $\sim
0.4$ mag for the brightest disks, declining to $\sim 0.3$ mag for
$\vcirc\approx 100\kms$. These values are close to those found by
\citet{tully98} for the extinction in a galaxy with average
inclination as a function of HI velocity width. The correction is
implemented {\it only} when comparing the observations with our model galaxies.

\subsection{Early-type galaxies}
\label{sec:app}

\begin{figure*}[htb!]
\epsscale{2.1}
\plottwo{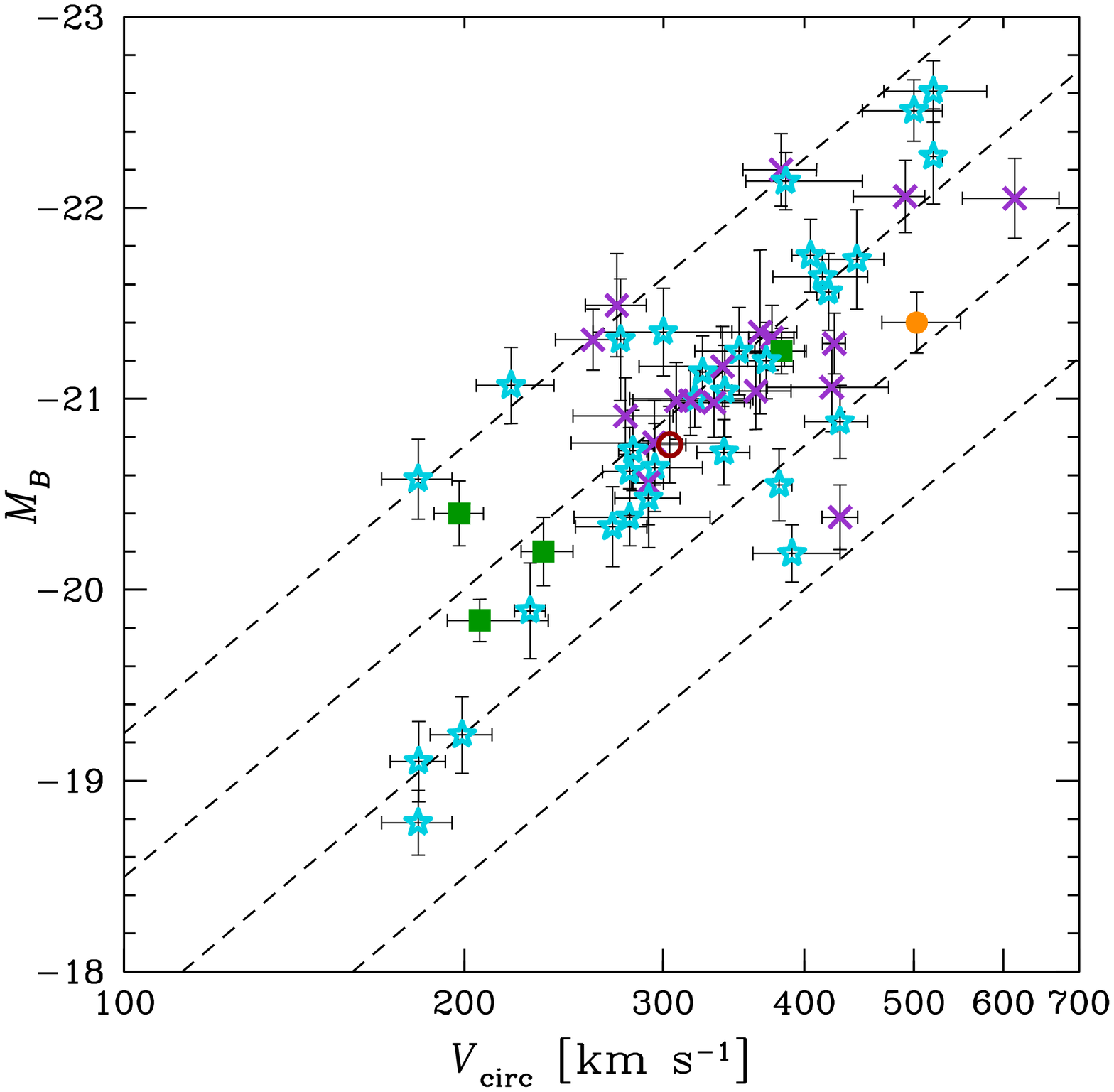}{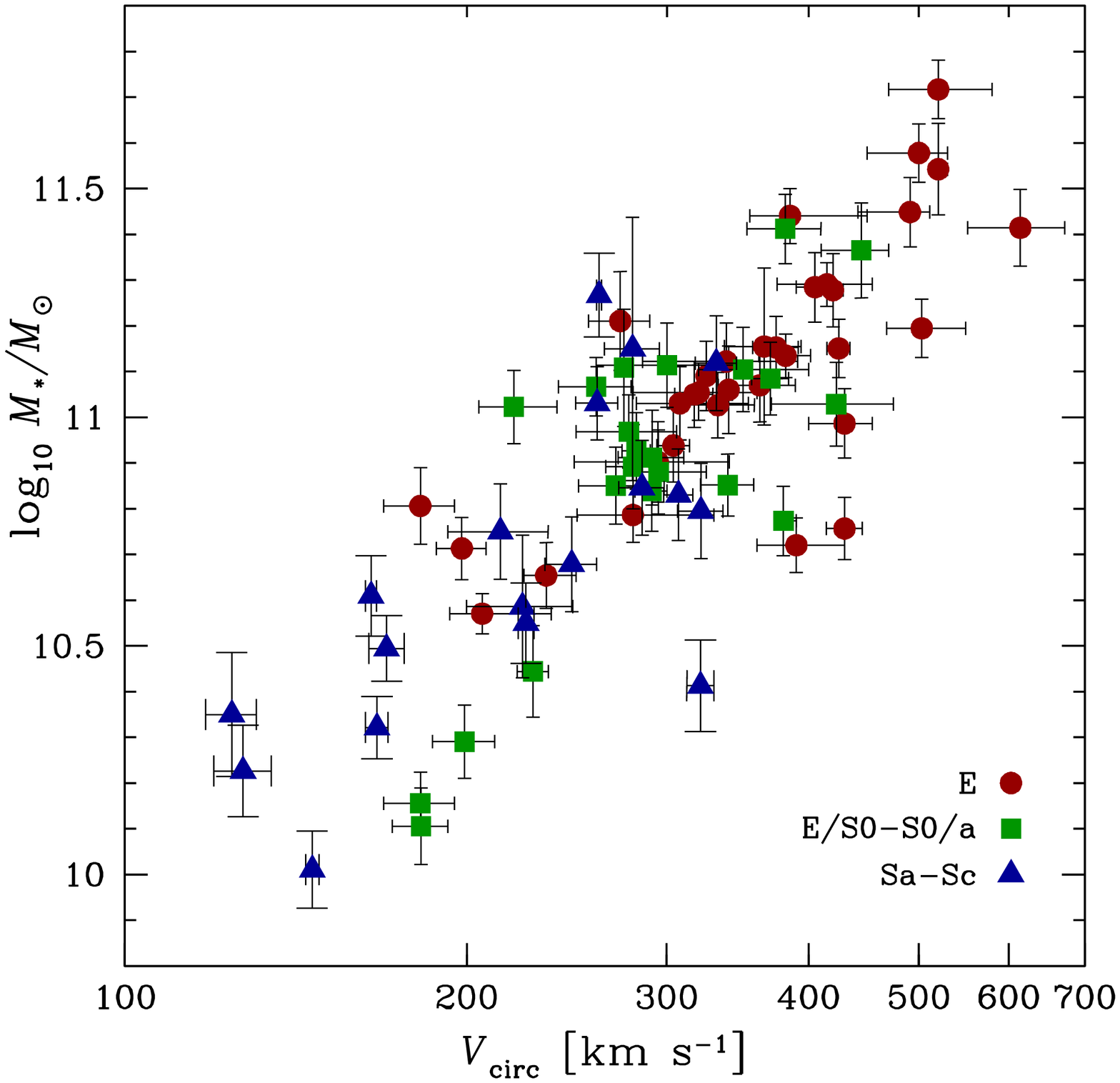}
\caption{
Properties of the early-type galaxy sample.  {\it Left panel:}
$B$-band luminosity versus circular velocity measured at 10 kpc for
individual galaxies. Symbols
indicate the mass probe used: stellar kinematics ({\it light blue
  stars}), X-ray gas ({\it purple crosses}), planetary nebula
kinematics ({\it green squares}), globular cluster kinematics ({\it
  orange filled circle}), and a cold gas ring ({\it red open
  circle}).  Dashed lines show $B$-band dynamical mass-to-light
ratios of $M/L_B=3$, 6, 12, and 24 ({\it top to bottom}) assuming all the light is contained
within 10 kpc; for
comparison, early-type galaxies are expected to have stellar
$M/L_B\sim 2.0$--$2.3$.  {\it Right panel:} Stellar mass as a function
of circular velocity at 10 kpc for ellipticals and S0s along with some late types shown
for reference.}  
\label{fig:ETG}
\end{figure*}

We also include bulge-dominated early-type galaxies
(ellipticals and lenticulars) in the LV relation, again measuring the
circular velocity at our fiducial 10~kpc radius.
The circular velocity
in this case is used not as a measure of rotation but merely as a
probe of the mass profile, further justifying the use of the term
 ``LV relation'' instead of TF relation.
 Using early-type galaxies
allows us to probe closer to the mass regime where the abundance of DM
halos drops exponentially (i.e., the knee of the velocity function),
which is more sensitive to the cosmological model. It also allows for study of halo-galaxy relations without
regard to the details of the evolution of the stellar populations
within them.

Because of the challenges of both observing and modeling early-type
galaxies, so far there exists no comprehensive set of mass
measurements for them akin to the spiral galaxy samples.  Instead, we
compile a set of high-quality LV estimates for individual galaxies
from the literature.

To provide the necessary LV data for nearby elliptical and lenticular
galaxies, we searched the literature for high-quality mass
measurements at $\sim 10 \kpc$ radii.  A variety of different mass
tracers were used including hot X-ray gas, a cold gas ring,
and kinematics of stars, globular clusters, and planetary nebulae.  We
required the mass models to incorporate spatially-resolved temperature
profiles in the case of X-ray studies, and to take some account of
orbital anisotropy effects in the case of dynamics.  We also used only
those cases where $V_{10}$ was constrained to better than $\sim$~15\%.
We make no pretense that this is a systematic, unbiased, or especially
accurate sample of early-type masses, noting simply that it is
preferable to ignoring completely this class of galaxies which
dominates the bright end of the luminosity function.

As an aside, we find in comparing to central velocity dispersions $\sigma_0$ taken
from HyperLeda \citep{Paturel03}, that the scaling $\vten \simeq
\sqrt{2} \sigma_0$ works very well on average, suggesting
near-isothermal density profiles over a wide range of galaxy masses.
It is far easier to measure $\sigma_0$ observationally than $\vten$,
motivating the use of the former as a proxy for the true $\vc$ which
is more robustly predicted by theory.  The $\sim$~15\% scatter that we
find in the $\sigma_0$--$\vten$ relation is relatively small, but it
is beyond the scope of this paper to consider the potential
systematics of using $\sigma_0$ as a proxy. We will use $\vten$ for the LV analysis in this paper.

For the luminosities, we make use of the total $B$-band apparent
magnitudes from the RC3 \citep{RC3}, corrected for Galactic
extinction. To correct to $^{0.1}r$ magnitudes, we use the
  filter conversions in \citet{Blanton07} together with $(B-V)$ colors
  obtained from HyperLeda \citep{Paturel03}.  For the distances (required
both for absolute magnitudes and for choosing the circular velocity
measurement radii in kpc), we use as a first choice the estimates from
surface brightness fluctuations \citep{Jensen03}, and otherwise the
recession velocity. 

The local data for 55 individual early-type galaxies are presented in
Figure~\ref{fig:ETG} (left). Dashed lines show $B$-band dynamical
mass-to-light ratios of $M/L_B=3$, 6, 12, and 24; for comparison,
early-type galaxies with typical colors ($B-V \sim 0.85$--$0.95$) are
expected to have stellar $M/L_B\sim 2.0$--$2.3$ for a Chabrier IMF
(e.g., Fig.~18 of \citet{Blanton07}). A table including the sources of
the data is provided in
Appendix~\ref{sec:ETGtable}. There is no obvious systematic difference
between the results from different mass tracers.  The galaxies appear
to trace a fairly tight LV sequence, except around the $L^*$
luminosity (assuming $M_B^* \approx -20.6$), where there are a few galaxies whose circular velocities appear to be fairly high or low.  The low-$V_{10}$ galaxies include
NGC~821 and NGC~4494, which were previously suggested as having a
``dearth of dark matter'' \citep{Romanowsky03}, and as implying a dark
matter ``gap'' with respect to X-ray bright ellipticals
\citep{Napolitano09}.  The present compilation suggests that the
galaxy population in the local universe may fill in this gap, although
further work will be needed to understand the scatter.

The right panel of Figure~\ref{fig:ETG} shows the relation between
stellar mass and circular velocity  at 10 kpc for the galaxies in the
early-type sample along with a few spirals for comparison. Stellar
masses were obtained using equation~(\ref{eq:MLratio}) as explained in
Section~\ref{sec:barfracBTF}. The
ellipticals are virtually indistinguishable from the S0s in the regime
where they overlap while the spirals seem to contain slightly more
stellar mass at the same $\vc$. We will discuss this issue in more
detail in Section~\ref{sec:barfracBTF} .

\subsection{Observational LV relation}

\begin{figure}[htb!]
\epsscale{1.00}
\plotone{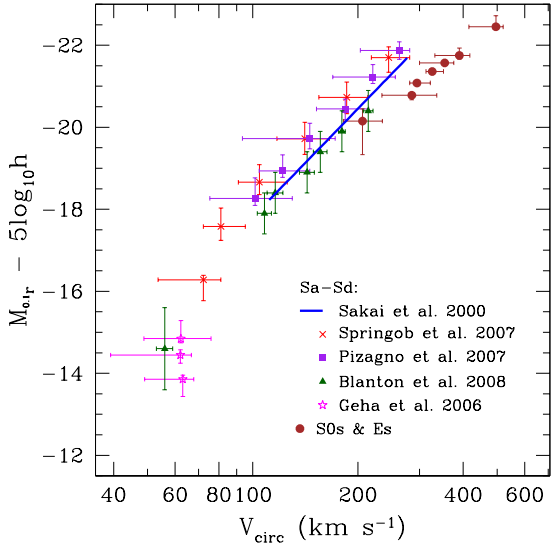}
\caption{The $r$-band luminosity versus circular velocity for several observational
samples across different morphological types.  All absolute magnitudes were
transformed to the SDSS $r$-band at redshift $z=0.1$. Points with
error bars show the median and 1-$\sigma$ scatter.}
\label{fig:LV0}
\end{figure}

Figure~\ref{fig:LV0} shows the combined LV relation for galaxies
with very different morphologies: from Magellanic dwarfs with $\vcirc\approx 50\kms$ to
giant ellipticals with $\vcirc\approx 500\kms$. The LV relation is not
a simple power-law. Dwarf galaxies show a tendency to have lower
luminosities as compared with a simple power-law extrapolation from
brighter magnitudes. There is a clear sign of bimodality at the bright
end of the LV relation with early type galaxies having $\sim 20 - 40\%$
larger circular velocities as compared with spiral galaxies with the same
$r$-band luminosity (or, conversely, $\sim 1$ magnitude fainter at fixed $\vc$). 

\section{The Bolshoi simulation}
\label{sec:sim}

The Bolshoi simulation was run using the following cosmological
parameters: $\Omega_{\rm matter}=0.27$, $h=0.70$, $\sigma_8=0.82$,
$\Omega_{\rm bar}=0.0469$, $n= 0.95$. These parameters are compatible
with the WMAP seven-year data (WMAP7) \citep{Jarosik10} and with WMAP5
combined with Baryon Acoustic Oscillations and Type 1a Supernova data
\citep{Hinshaw09, komatsu09,Dunkley09}. The Bolshoi parameters are in
excellent agreement with the SDSS maxBCG+WMAP5 cosmological parameters
\citep{Rozo09} and with cosmological parameters from WMAP5 plus recent 
X-ray cluster survey results \citep{klypin10}.

It is important to appreciate that Bolshoi differs essentially 
from another large, DM-only cosmological simulation: the Millennium
simulation \citep[][MS-I]{MSI}. The Millennium simulation has
been the basis for many studies of the distribution and statistical
properties of dark matter halos and for semi-analytic models of the
evolving galaxy population. This simulation and the more recent
Millennium-II simulation \citep[][MS-II]{MSII} used the first-year (WMAP1) cosmological
parameters, which are rather different from the most recent
estimates. The main difference is that the Millennium simulations used
a substantially larger amplitude of perturbations than
Bolshoi. Formally, the value of $\sigma_8$ used in the Millennium
simulations is more than 3$\sigma$ away from the WMAP5+BAO+SN value
and nearly 4$\sigma$ away from the WMAP7+BAO+H$_0$ value.  However,
the difference is even larger on galaxy scales because the Millennium
simulations also used a larger tilt $n=1$ of the power spectrum. 

The Bolshoi simulation uses a computational box $250\Mpch$ across and
$2048^3\approx$~8.6 billion particles, which gives a mass resolution
(one particle mass) of $m_1=1.35\times 10^8\Msunh$. The force
resolution (smallest cell size) is physical (proper) 1~\kpch. For
comparison, the Millennium-I simulation had a force resolution (Plummer
softening length) 5~\kpch~ and the Millennium-II simulation had
1~\kpch. 
The Bolshoi simulation was run with the Adaptive-Refinement-Tree (ART)
code, which is an Adaptive-Mesh-Refinement (AMR) type code. A detailed
description of the code is given in \citet{Kravtsov97,Kravtsov99}.
We refer the reader to 
\citet{klypin10} for more details specific to the use of the code for
the simulation.

We use a parallel (MPI+OpenMP) version of the Bound-Density-Maxima (BDM)
algorithm to identify halos in Bolshoi \citep{Klypin97}. BDM does not
distinguish between halos and subhalos\footnote{A subhalo is a halo
  which resides within the virial radius of a larger halo.} -- they are treated in the same
way. The code locates maxima of density in the distribution of
particles, removes unbound particles, and provides several statistics
for halos including virial mass and radius, as well as density
profiles\footnote{The Bolshoi halo catalogs are
  publicly available at {\url {http://www.multidark.org}}.}.
We use the virial mass definition $\Mvir$ that follows from the
top-hat model in the expanding Universe with a cosmological
constant. We define the virial radius $\Rvir$ of  halos as the radius
within which the mean density is the virial overdensity times the mean
universal matter density $\rho_{\rm m}=\Omega_{\rm m}\rho_{\rm crit}$ at that
redshift. Thus, the virial mass is given by
\begin{equation}
    M_{\rm vir} \equiv {{4 \pi} \over 3} \Delta_{\rm vir} \rho_{\rm m} R_{\rm vir}^3 \ .
\end{equation}
For our set of cosmological parameters, at $z=0$ the virial radius
$\Rvir$ is defined as the radius of a sphere enclosing average overdensity equal to
$ \Delta_{\rm vir}= 360$ times the average matter density. The overdensity limit changes
with redshift and asymptotically goes to 178 for high $z$.  Different
definitions are also found in the literature. For example, the often
used overdensity 200 relative to the {\it critical} density gives mass
$M_{200}$, which for Milky-Way-mass halos is about 1.2-1.3 times
smaller than $\Mvir$. The exact relation depends on halo
concentration.

At each timestep there are about 10~million halos in Bolshoi (8.8$\times10^6$
at $z=0$, 12.3~$\times10^6$ at $z=2$, 4.8~$\times10^6$ at $z=5$). The
halo catalogs are complete for halos with $\vc>50$~\kms~ ($\Mvir
\approx 1.5\times 10^{10}\Msunh$).  In order to track evolution of
halos over time, we find and store the 50 most bound particles.
Together with other parameters of the halo (coordinates, velocities,
virial mass, and circular velocity) the information on most bound
particles is used to identify the same halos at different moments of
time.  The procedure of halo tracking starts at $z=0$ and goes back in
time. The final result is the history ({\it track}) of the major
progenitor of a given halo.

\section{DM halos: definitions and characteristics}
\label{sec:halo}

We distinguish between two types of halos. A halo can be either distinct (not
inside the virial radius of a larger halo), or a subhalo if it is inside of a larger halo.
For both distinct halos and subhalos, the BDM halo finder
provides the maximum circular velocity
\begin{equation}
 \vc = \sqrt{\frac{GM(<r)}{r}}\Big|_{\rm max} .
\end{equation}
Throughout this paper we will use term {\it circular velocity} to mean
maximum circular velocity. 

As  the main characteristic of the DM
halos we use their circular velocity $\vc$. There are advantages to using
$\vc$ as compared with the virial mass $\Mvir$. The virial mass is a
well defined quantity for distinct halos, but it is ambiguous for
subhalos. It strongly depends on how a particular halo-finder code
defines the truncation radius and removes unbound
particles. It also depends on the distance to the center of the host
halo because of the effects of tidal stripping. Instead, the circular
velocity is less prone to those complications.  The main motivation
for using $\vc$ in this work is that it is more closely related to the
properties of the central regions of halos and, thus, to galaxies
hosted by those halos.  For example, for a Milky-Way type halo the
radius of the maximum circular velocity is about 40~kpc (and $\vc$ is
nearly the same at 20~kpc), while the virial radius is about
200~kpc. In addition, the virial mass of a DM halo is not an easily
observable quantity and this further limits its use for comparison of
simulations with observations.

Tidal stripping can lead to significant mass loss in the
periphery of subhalos. The net effect at redshift zero of the complex
interactions that each halo undergoes is a decrease in the
maximum circular velocity compared to its peak value over the entire
history of the halo. The galaxy residing in the central region of the
halo should not experience much stripping and should preserve
most of its mass inside the optical radius
\citep[e.g.,][]{nagai05,conroy06}. Following this argument, the initial total
mass distribution and rotation profile of the halo are frozen at the
moment before the halo is accreted and starts to experience stripping. We refer to this circular velocity as $\vacc$.  In practice
we find the peak circular velocity of the halo over its entire
history. Further details on the halo tracking procedure can be found
in \citet{klypin10}.

\section{Connecting galaxies and DM halos}
\label{sec:procedure}

To investigate the statistics of galaxies and their relation to host
DM halos as predicted by the $\LCDM$ model using HAM, 
we obtained the properties of our model galaxies using the following procedure:

\begin{enumerate}

\item Using the merger tree of each DM halo and subhalo,
  obtain $\vacc=$  the peak value of the circular velocity over the history of
  the halo (this is typically the circular velocity of the halo
  when it is first accreted).  Perform abundance matching of the velocity
  function of the  halos to the LF of galaxies to obtain the
  luminosity of each model galaxy.  

\item Perform abundance matching of the velocity function to the
  stellar mass function of galaxies to obtain the stellar mass of each model
  galaxy.

\item Use the observed gas-to-stellar mass ratio as a function
  of stellar mass to assign cold gas
  masses to our model galaxies. The stellar mass added to the cold gas
  mass becomes the cold baryonic mass.

\item Using the density profiles of the DM halos, obtain the circular
  velocity at $10\kpc$ ($\vten$) from the center of each
  halo. To do this, calculate the dark-matter-only contribution by multiplying the DM mass profile, obtained directly from the simulation, by the
  factor $(1-f_{\rm bar})$, where $f_{\rm bar}$ is the cosmological
  fraction of baryons\footnote{Recall 
  that the Bolshoi simulation was run for a dissipationless cosmic density
  $\Omega_{\rm m}=\Omega_{\rm dm} + \Omega_{\rm bar} = 0.27 = 
  \Omega_{\rm dm} (1 + f_{\rm bar})$.}.
  Then take the total cold baryon contribution from step 3 and assume it is
  enclosed within a radius of $10\kpc$. Adding the two contributions gives the total mass required to calculate $\vten$.

\item Implement the correction to $\vten$ due to the adiabatic
  contraction of the DM halos due to the infall of the cold baryon component to the center.

\end{enumerate}

We now explain each of the above 5 steps in detail.

Using the key assumption that halos with deeper potential wells become
sites where more baryons can gather to form larger and more luminous
galaxies, we ranked our halos and subhalos using their $\vc$, and starting from the
bright end of the $r$-band LF, assigned luminosities to each according to their
space density using the prescription found in \citet{conroy06}. In
other words, we found the unique one-to-one correspondence that would
match the halo velocity function with the luminosity function of
observed galaxies. Of course, this is a simplifying approximation.  It
does not discriminate between blue (star-forming) and red (quenched) galaxies, for example.

In this paper we use the Schechter fit to the $r$-band
galaxy LF measurement of \citet{montero-dorta09} obtained from the
SDSS Data Release 6 (DR6) galaxy sample\footnote{To avoid aperture
corrections when comparing to other data we use model magnitudes instead of
Petrosian values.}. The fit is characterized by the parameters:
$\Phi^{\ast} = 0.0078$, $M_{0.1r}^* - 5\log h= -20.83$, and $\alpha
=-1.24$. Since the median redshift of
the SDSS DR6 sample is $\approx 0.1$ \citep{blanton03a}, all our
subsequent results will be shown in $^{0.1}r$-band magnitudes.  

As an alternative, we also consider a LF with a steeper slope at
low-luminosities.  \citet{blanton05} obtained the SDSS LF including
dwarfs as faint as $M_r=-12$ and investigated surface brightness
completeness at the faint end of the distribution. Their steeper value
of the faint-end slope was obtained by  weighting the abundance of
each galaxy by its estimated surface brightness completeness. To quantify the effect of
including low surface brightness galaxies in our model (those with
Petrosian half-light $r$-band surface brightness greater than 24.0 mag/arcsec$^2$),
we increased the abundance of bright galaxies in the \citet{blanton05}
LF to match the \citet{montero-dorta09} LF at the bright end while
keeping the steep faint-end slope ($\alpha = -1.34$) measured by
\citet{blanton05}. The modified LF produces 60\% more
galaxies at $M_r \sim -16$ and over a factor of 2.5 more galaxies at $M_r
\sim -13$. Using this LF to perform the abundance matching increases
the luminosity of galaxies assigned to small DM halos, steepening the
faint end of the LV relation.

It is important to note
that we assume that each dark matter halo or subhalo must contain a
galaxy with a detectable luminous component (for the SDSS $r$-band
this requires galaxies to be detectable in visible wavelengths) and
this component must evolve in a way that guarantees its detectability
at $z=0$.  Since the effective volume surveyed by SDSS DR6 at $z < 0.3$
is comparable to the volume of the Bolshoi simulation, we expect the
statistics of the halo population to be comparable to those of the
observed galaxies all the way up to the large mass/luminosity tail of
the distributions.  Even though Bolshoi contains a factor of $\sim
1.8$ more objects than the sample of \citet{montero-dorta09},
abundance matching is mostly insensitive to uncertainties in the 
high-luminosity tail of the LF.

\subsection{The circular velocity of galaxies inside 
halos including cold baryons}
\label{sec:baryons2}

The next step is to separate the DM and baryon components in each halo
and allow the baryons to dissipatively sink to the centers of the DM
halos. We assume 
for simplicity that there is a radius at which we could consider
most of the cold baryons to be enclosed, with
only dark matter present beyond that point. 

This cold baryon component has been observed to comprise only a small fraction of the cosmic 
abundance of baryons; in other words, the cold baryon fraction 
$\fbar \equiv \Mbar/(\MDM + \Mbar)$ in galaxies is much lower 
\citep{fukugita98,fukugitapeebles04}
than $\Omegabar/ \Omegamatter = 0.17$ \citep{komatsu09}. We resort to the observations and use the galaxy stellar mass function
obtained from the SDSS DR7 by \citet{li09}, who employ estimates of 
stellar masses by \citet{Blanton07} obtained using five-band SDSS
photometry assuming the universal IMF of \citet{chabrier03}. These
masses are consistent with those estimated using single-color and
spectroscopic techniques \citep{li09}. 

Using the same procedure described above for the luminosity function,
we abundance-matched the halos in Bolshoi to the galaxies in the SDSS DR7
starting from the high stellar mass end until reaching our
completeness limit at $\vc=50\kms$, obtaining stellar masses for
each galaxy. Strictly speaking, this procedure results in a one-to-one
relation between circular velocity and stellar mass-to-light ratio
which should only be interpreted as the {\it average} of a population of
galaxies with a given $\vcirc$. The scatter (or bimodality) in the
mass-to-light ratio as a function of circular velocity could be measured from observations and included
in the assignment but would not change our results significantly. 

Since dwarfs can have most of their cold baryons in the
gas phase instead of in stars \citep{baldry08}, we calculated for each model
galaxy the total cold gas mass using a parameterization of the
observed atomic gas mass fraction as a function of stellar mass from
\citet{baldry08} (their equation (9), shown as a dashed line in their
Figure 11). This includes the total cold atomic gas found in the disk only. The gas-to-stellar mass ratio $f_{\rm
  gas}$ depends on stellar mass and it is the largest for dwarfs. For example,
$f_{\rm gas}\approx 4-5$ for galaxies with $M_*=10^8\Msun$. It
declines to $f_{\rm gas}\approx 0.25 (0.1)$ for galaxies with
$M_*=10^{10} (10^{11})\Msun$. It should be even smaller for
ellipticals and S0s. It should be noted that, when it comes to
dynamical corrections to
$\vcirc$, the gas plays a minor role. It only becomes important when we
consider the baryonic Tully-Fisher relation.

Lastly, we add the stellar and cold gas masses for each model galaxy
and obtain the correction to the circular velocity of the pure DM
halo at a radius enclosing the cold baryonic mass. We set this value
to $10\kpc$ for all the halos in our sample. In the case of dwarf
galaxies this should  be a good approximation to the maximum circular
velocity since their rotation curves rise much more slowly and in some
cases they peak beyond the optical radius \citep{courteau97}. Our
assumption allows us to include the peak of the rotation curve for
most of these galaxies. In the case that the peak is located well
within $10\kpc$ the correction would be almost negligible since we
would be still measuring rotation in the flat regime. Additionally,
truncating the cold baryons at a radius of $10\kpc$ allows us to directly
calculate the correction to the circular velocity at that radius
without having to resort to more complicated assumptions about the
distribution of baryonic matter in galaxies, i.e., exponential lengths
and S\'ersic indices of disks and bulges as well as extended gas and stellar halos.

To obtain the circular velocity measured at $10\kpc$ ($\vten$) for the
Bolshoi DM halos, we need to use dark matter profiles and find the
dark matter mass inside a 10~kpc radius. To do this, we could use the
individual profile of each halo. However, once we select halos with
a given maximum circular velocity, individual halo-to-halo variations
are small at 10~kpc (the situation is different if we select halos
using virial mass, which results in large deviations in concentration producing large variations in $\vten$). This is why instead of individual
profiles we construct average profiles for halos within a narrow range
$\Delta \log_{10}V \approx 0.05$ of maximum circular velocity.

\begin{figure}[htb!]
\epsscale{1.0}
\plotone{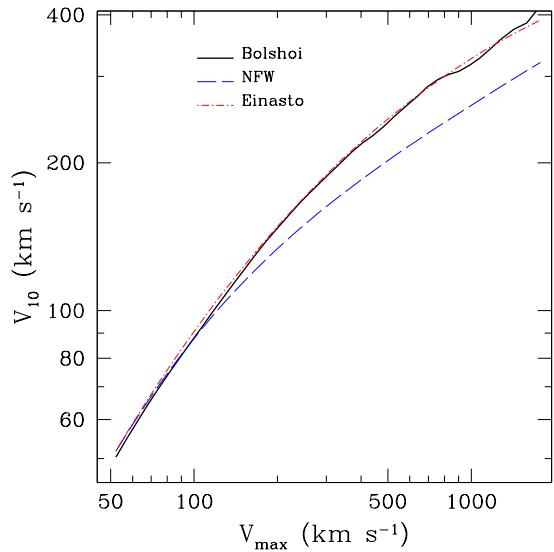}
\caption{Relation between the maximum circular velocity ($\vmax$) and the circular
  velocity measured at $10\kpc$ ($\vten$) for the dark matter halo only
  (excluding the cold baryonic component). The solid curve shows the binned
  median values of the Bolshoi DM halo sample. The other curves show
  the relation obtained assuming the NFW (dashed) and the Einasto
  (dot-dashed) profiles with the halo concentration given by
  eq. (\ref{eq:conc}).}
\label{fig:V10}
\end{figure}

We first bin and average the circular velocity profiles of the distinct halos
found by the BDM code. These profiles are calculated for each halo
(including unbound particles) in logarithmic radial bins in units of
$\Rvir$. Using distinct halos is convenient because it gives us
density profiles that are less affected by interactions than those of
subhalos. For the inner profiles of subhalos the effect is relatively small because the
stripping happens preferentially at the outer radii. Using the
averaged binned circular velocity profiles we obtain the  velocity at
$10\kpc$. Within about $1.2\%$ of the virial radius, discreteness
effects render the profiles unreliable and we use instead
extrapolation with the shape of a simple power-law in radius.  For
halos with $\vc <100\kms$ the full extent of the profiles suffer from
measurement noise which we avoid by extrapolating from the profile of
halos with $\sim 100 \kms$. Figure~\ref{fig:V10} shows the obtained
median relation between the maximum circular velocity $V_{\rm max}$ and $\vten$ for the Bolshoi DM halos
without inclusion of the cold baryons.

The estimates of the relations obtained when a parametric form of the
density profile is used are also shown for the case of the NFW
\citep{navarro97}
\begin{equation}
	\rho(r) = \frac{4\rho_s}{(r/\rs)(1+r/\rs)^2} ,  
\label{eq:NFW}
\end{equation}
and the  Einasto \citep{einasto65,graham06} universal profiles
\begin{equation}
	\rho(r) = \rho_s \exp \left\{ -2n \left[(r/\rs)^{1/n} - 1 \right] \right\} ; 
\label{eq:Einasto}
\end{equation}
where $\rs$ is the radius at which the logarithmic slope of the
density profile is $-2$. Following \citet{graham06}, we use $n=6.0$.
The concentration parameter defined for both models as $c \equiv
\Rvir/\rs$ is given by the relations obtained in Paper I for distinct
halos (\citealt{klypin10}; see also \citealt{prada11}) :
\begin{equation}
	c = 9.60 \left(\frac{\Mvir}{10^{12} \Msunh} \right)^{-0.075}, 
\label{eq:conc}
\end{equation}
and
\begin{equation} 
	\Mvir = \left( \frac{\vc}{ 2.8 \times 10^{-2} \kms} \right)^{3.16} \Msunh.   
\label{eq:conc2}
\end{equation}
Note that in Figure~\ref{fig:V10} we use total dynamical masses and do not
account for the condensation of baryons. For $\vc =100-450 \kms$ the rotation (or density)
profiles of the Bolshoi simulation are extremely well approximated by
the Einasto parameterization, whereas NFW underestimates $\vten$ by
almost $20\%$ at $450\kms$. Following the conclusions of
\citet{navarro04} and \citet{graham06}, we assume the Einasto profile
when extrapolating the inner parts of the largest ($\vcirc>450\kms$) halos.

\begin{figure*}[htb!]
\epsscale{1.50}
\plotone{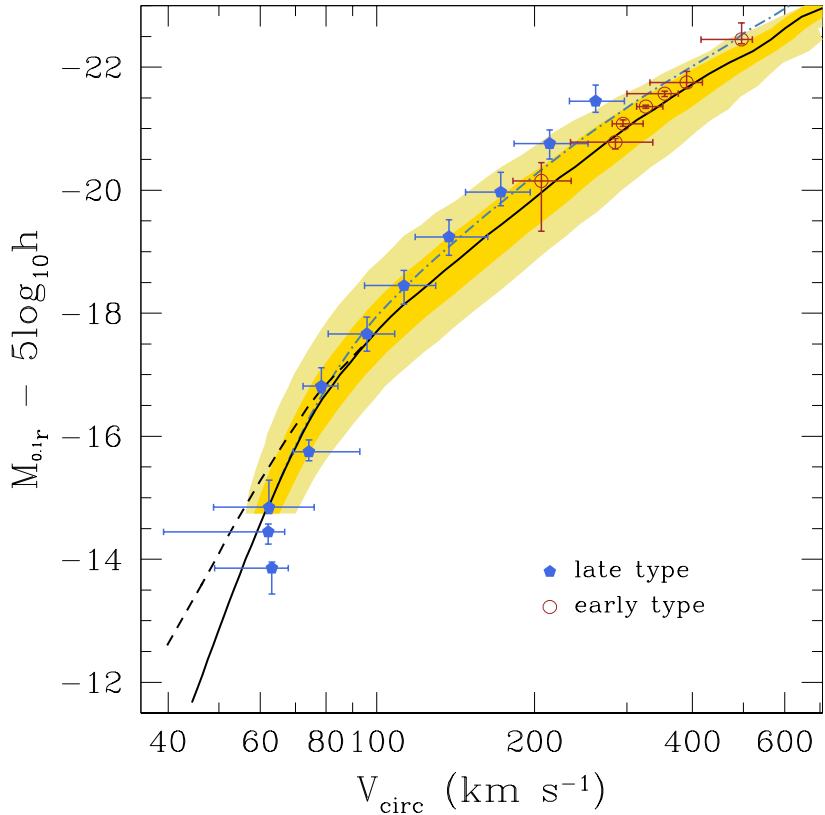}
\caption{Comparison of the observed Luminosity-Velocity relation with
  the predictions of the $\LCDM$ model using halo abundance matching. The solid curve shows the median
  values of $^{0.1}r$-band luminosity vs. circular velocity for the model
  galaxy sample. The shaded regions enclose 68\% and 95\% of the model
  galaxies in bins of luminosity. The circular velocity for each model galaxy is based
  on the peak circular velocity of its host halo over its entire
  history, measured at a distance of 10~kpc from the center including
  the cold baryonic mass and the standard correction due to adiabatic
  halo contraction. The dashed curve shows results for a steeper 
  ($\alpha = -1.34$) slope of the LF. The dot-dashed curve shows predictions after
  adding the cold baryon mass but without adiabatic halo contraction. Points show
  median and 1-$\sigma$ scatter of representative observational samples.}
\label{fig:LV1}
\end{figure*}

\subsection{Adiabatic contraction of DM halos}
\label{sec:AC2}

Dissipation allows the baryons to condense into  galaxies in the
central regions of DM halos dragging the surrounding dark matter into
a new more concentrated equilibrium configuration. If the density of
the DM halo increases considerably within the extent of the disk, the
peak circular velocity could be much larger than our previous
estimates. There are different approximations for the adiabatic
compression of the dark matter. \citet{blumenthal86} provide a simple
analytical expression, which is known to overpredict the
effect. The approximation proposed by \citet{gnedin04} predicts
significantly smaller increase in the density of the dark matter. More
recent simulations indicate even smaller compression
\citep{Tissera09,Duffy10}. However, at a 10~kpc radius the dark matter
contributes a relatively large fraction of mass even for large
galaxies. As a result, the difference between the strong
compression model of \citet{blumenthal86} and no-compression is only
$10-20$\% in velocity. 

We use the standard adiabatic contraction (AC) model of
\citet{blumenthal86} to bracket the possible effect.
We thus assume that following the condensation of the
baryons, the dark matter particles adjust the radius of their orbits
while conserving angular momentum in the process. We solve the
equation
\begin{equation}
	\Mtot(r_{\rm i})r_{\rm i} = [\MDM(r_{\rm i})(1-f_{\rm bar}) + \Mbar(r_{\rm f})]r_{\rm f} , 
 \label{eq:AC}
\end{equation}
where $r_{\rm f} = 10\kpc$, $\Mbar(r_{\rm f})$ is the total baryonic mass assigned
to each halo and $f_{\rm bar}=\Omega_{\rm bar}/\Omega_{\rm matter}$ is the
universal fraction of baryons. We solve  equation~(\ref{eq:AC}) for
$r_{\rm i}$ and then add the dark matter mass $\MDM(r_{\rm i})(1-f_{\rm bar})$ to
the mass of cold baryons to find the circular velocity. Note that this
implies that only cold baryons (i.e., stars and cold gas) are left in
the central regions of the galaxy, while the remaining hot baryons 
are at larger radii.  As expected, only the halos that
are dominated by baryons at their centers suffer a significant
increase in their measured circular velocities due to the increase
in concentration of dark matter as result of adiabatic contraction.

\section{Results}
\label{sec:results}

\subsection{The Luminosity-Velocity relation}
\label{sec:TF}

Figure~\ref{fig:LV1} shows the predicted LV relation for galaxies in
the $\LCDM$ model obtained using halo abundance matching. We binned
together the major observational samples described in
Section~\ref{sec:observations} and include them for comparison.  The
internal extinction de-correction for late-type galaxies described in Section~\ref{sec:LTG} was
implemented for a fair comparison with the models. The full curve in
this plot shows results obtained using the \citet{montero-dorta09} LF of galaxies in the SDSS
DR6 sample and uses the adiabatic contraction model of
\citet{blumenthal86}. The 1- and 2-$\sigma$ width of the distribution
of model galaxies in bins of luminosity is represented by the shaded
regions (details about the model used to add scatter can be found in Section~\ref{sec:scatter}). Predictions without adiabatic contraction
(with the cold baryon contribution added in quadrature) are shown as the
dot-dashed curve. The dashed curve shows the effect of a steeper slope
at the faint end of the LF that accounts for potential surface brightness
incompleteness.  (For details see Section~\ref{sec:procedure}).
Although dwarf galaxies with $\vc < 80\kms$ seem to agree
better with a model using the \citet{montero-dorta09} luminosity function, the scatter
of the observed dwarf LV relation is so large that both LFs used in
conjunction with the abundance of DM halos produce results that are
consistent with the available data.

One may note that the AC model misses late-type points with
$\vcirc=150-250\kms$ and the no-AC model practically fits most
of the late-type measurements. This should {\it not} be
interpreted as either an advantage for the no-AC model or a
disadvantage for the AC model. Our predictions apply to the average
population across all types of galaxies. Because of the
dichotomy of the LV diagram, a model that goes through either early-type galaxies or through late-type galaxies is an incorrect model. The
correct answer should be a model which tends to be close to spirals at low
luminosities (where spirals dominate the statistics) and gradually slides towards the
ellipticals at the high-luminosity tail where they dominate. It
seems that the AC model does exactly that, but it may overpredict the
circular velocities at
the very bright tail of the LV relation.\footnote{\citet{Schulz10},
  \citet{Napolitano10} and \citet{Tortora10} find observational
  evidence for AC in elliptical galaxies.}

As demonstrated in Section~\ref{sec:scatter}, our stochastic HAM model
accounts for galaxies that reside in halos with both smaller and
larger $\vc$ than the average. For example, since the most massive spiral galaxies comprise a very small percentage of the
galaxy population with $\vc > 250\kms$ (less than 10\%),
in our model they are assigned to the low-$\vc$ wing of the distribution shown in
Figure~\ref{fig:LV1}. Hence, even though the model makes predictions
for the average galaxy population, it also accounts for the
morphological bimodality observed in the LV relation.

One also should not overestimate the quality of the observations. The
fact that in Figure~\ref{fig:LV1} the brightest spirals with $M_r - 5\log_{10} \lesssim -21$ are more
than 2-$\sigma$ away from the mean of the models is not a problem because of the size of the uncertainties in the
observational data. For instance, there is a systematic $\sim 10$\%
velocity offset between the measurements of \citet{pizagno07} and
\citet{springob07}, which seems to point to the current uncertainties
in the LV relation.

Considering the current level of the uncertainties, our model galaxies
show remarkable agreement with observations spanning an order of
magnitude in circular velocity (or, equivalently, three orders of
magnitude in halo mass) and more than three orders of magnitude
in luminosity. For galaxies above $200\kms$, our model produces
results that agree extremely well with the observed luminosities of
early-types (Es and S0s).  Given how simple the prescription is, it is
perhaps surprising how closely we can reproduce the properties of observed
 galaxies. For galaxies with $\vcirc = 100-200\kms$,
DM halos without any corrections already occupy the region expected
for galaxies. The dynamical corrections improve the fits, but it is important to
emphasize that the abundance matching method yields the correct
normalization of the LV relation regardless of the details of the
corrections we implement.  Another feature of the relation, its
steepening below $100 \kms$, is caused by the the shallow faint-end slope of the
luminosity function. Although our model galaxies in this regime are slightly
underluminous as compared with a simple power-law extrapolation from
brighter galaxies, observed dwarfs seem to predict a deviation from a
power-law TF in the same general direction.

In the way it was constructed, our model galaxy sample does not include
uncertainties in either the halo velocity function or in the galaxy
luminosity function. This produces an LV relation with no
scatter. Section~\ref{sec:scatter} examines the effects of including
scatter in the halo matching procedure.

We now discuss in greater detail the results of the
individual steps explained in Section~\ref{sec:procedure}.

\subsection{The Luminosity-Velocity relation: detailed analysis of model components}

\subsubsection{Measuring circular velocity at 10~kpc}

\begin{figure}[htb!]
\epsscale{1.2}
\plotone{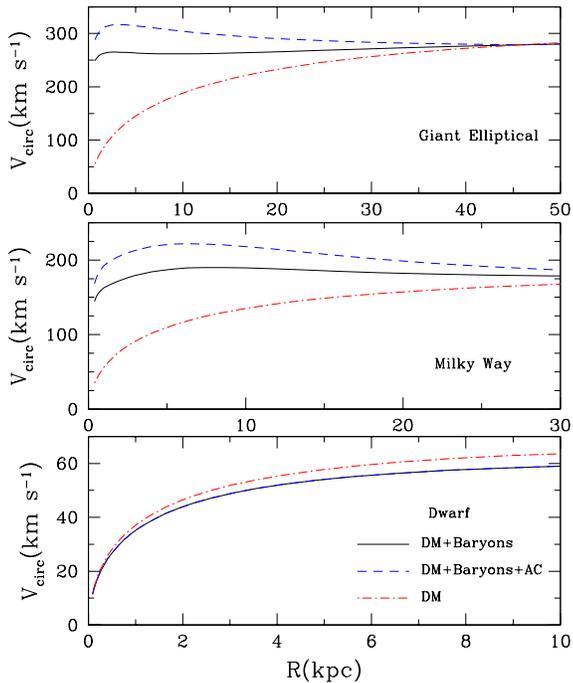}
\caption{Effect of cold baryons on circular velocity profiles for three
  characteristic models of galaxies with virial masses $10^{13}\Msun$
  (top), $1.7\times 10^{12}\Msun$ (middle), and $7\times 10^{10}\Msun$
  (bottom). The ``DM'' curves include a cosmological fraction of
  baryons that trace the dark matter distribution. The cold baryon
  mass is added to the true dark matter mass in calculating
  the circular velocity (``DM+Baryons''). The effect of adiabatic compression
  of the dark matter is included in the models named ``DM+Baryons+AC''.  After
  adding the cold baryons the circular velocities are rather flat in the
  inner $5-10$~kpc regions.}
\label{fig:Vcirc}
\end{figure}

Observations do not always provide measurements of the circular velocity
at 10~kpc.  This is especially true for dwarf galaxies where the last
measured point of the rotation curve can be at 3-5~kpc. What are the errors associated with
using measurements at different radii? Figure~\ref{fig:Vcirc} presents
three typical examples of the circular velocities expected for
galaxies with vastly different masses.

The top panel shows a model of a giant elliptical galaxy with $1.5\times
10^{11}\Msun$ of stellar mass distributed according to a $R^{1/4}$ law
with a half-light radius of $5.5$~kpc. The stellar component is
embedded in a dark matter halo with virial mass $\Mvir = 10^{13}\Msun$
and median concentration $c=7$. The maximum circular velocity
(310~\kms) of the halo is reached at 160~kpc.  The middle panel shows
a Milky Way-size model with maximum circular velocity 190~\kms, virial mass $\Mvir =
1.7\times 10^{12}\Msun$, and
median concentration $c=9$ for its mass. The cold baryonic component
consists of a Hernquist bulge ($M_{\rm bulge} =10^{10}\Msun$,
half-mass radius $R_{\rm bulge} =1$~kpc) and an exponential disk
($M_{\rm disk} =5\times 10^{10}\Msun$, scale radius $R_{\rm disk}
=3.5$~kpc). The bottom panel shows a dwarf galaxy model with $\Mvir =
7\times 10^{10}\Msun$, $c=12$, $V_{\rm max}=65$~\kms. Its cold baryons have two
exponential disks: one for stars and and one for cold gas with a mass
ratio of 1:4 and radii $R_*=1.5$~kpc and $R_{\rm gas}=3.0$~kpc. The
total mass in cold baryons is $M_{\rm bar}=3\times 10^8\Msun$. When we
include baryons, we assume that most of them were blown out
from the models and the only baryons left are in the form of stars and
cold gas. As in Bolshoi, the ``DM'' circular velocities in
Figure~\ref{fig:Vcirc} include a cosmological
amount of baryons that traces the distribution of the dark
matter. This contribution is removed from the mass profiles before
adding the cold baryons.  In all three cases we use the
Einasto dark matter profiles (equation~(\ref{eq:Einasto}))  with $n=6$. For the models labeled ``DM+Baryons'' at each
radius we simply add the mass of cold baryons and the mass of dark
matter. The models termed ``DM+Baryons+AC'' include adiabatic
compression of dark matter according to the prescription of \citet{blumenthal86}.

After adding the cold baryons the circular velocity profiles become rather
flat in the inner $5-10$~kpc regions of the galaxies implying that
measurements of circular velocities anywhere in this region are
accurate enough to provide the value 
of the circular velocity at 10~kpc.

There are some caveats in choosing 10~kpc as a fiducial radius
for either extremely massive spirals or giant
ellipticals.  Considering the correlation between central surface
brightness and disk scale-length found by \citet{courteau07}, the most
luminous disks appear to have scale-lengths as large as
$15\kpc$, whereas according to \citet{courteau97} their
rotation curves peak at about 1 scale-length. Hence, we may be {\it
  underestimating} the maximum observed rotation velocity of these
galaxies in our sample. We may also {\it overestimate} circular velocities when we assume that
most of the cold baryon mass is inside 10~kpc radius. In
principle, some corrections can be applied to compensate this
effect. However, our estimates show that at most this is a $\sim 20$\%
effect for spirals and somewhat smaller for ellipticals because they
are more compact for the same luminosity. Considering existing
uncertainties and complexities of implementing the correction, we
decided not to use them.

\subsubsection{Dark matter profiles}
\label{sec:accretion}

\begin{figure}[htb!]
\epsscale{1.0}
\plotone{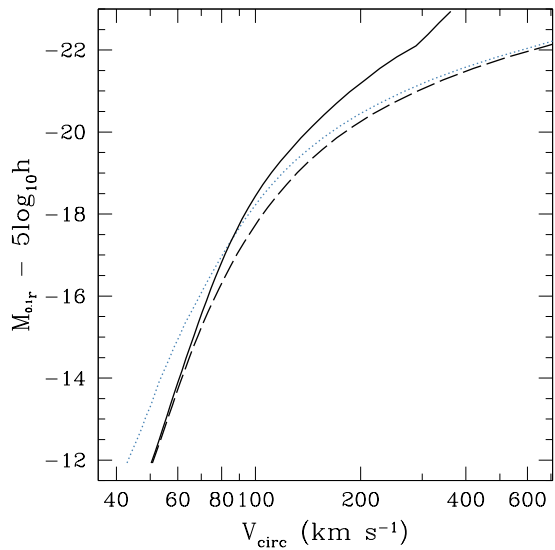}
\caption{Comparison between different effects on the measured circular
  velocities of  model galaxies without corrections for cold baryons or adiabatic contraction.  The
  dotted curve shows the median $^{0.1}r$-band luminosity vs. circular
  velocity of the model galaxies that results from abundance matching
  using the maximum circular velocity of each DM halo at $z=0$. The dashed line shows
  the effect of using the peak value of the maximum circular velocity {\it
    over the history of each halo} ($\vacc$). The solid curve shows
  the result of measuring $\vacc$ for each halo at $10\kpc$ from the
  center. This affects intermediate and large halos the most since their
  circular velocity profile is still rising at that distance. All the
  curves include a cosmic baryon contribution that traces the dark matter. }
\label{fig:LV2}
\end{figure}

To illustrate the effect of tidal stripping, Figure~\ref{fig:LV2} shows
the results (dashed curve) obtained when the luminosity assignment is
performed using the peak historical value of each halo's circular
velocity ($V_{\rm acc}$) as compared with the circular velocity
at $z=0$ (dotted curve). The reason why the dashed curve is rightwards of the
dotted one is that for subhalos the circular velocity estimated at
$z=0$ is smaller than its value at accretion $V_{\rm acc}$. If we
compare luminosities at the same circular velocity, the differences can be substantial: almost one magnitude for galaxies
with $V_{\rm circ} =50-60$~\kms due to the steep slope of the
LV relation for dwarfs. In terms of velocities, the differences are much
smaller.  Neglecting the effects of stripping in the assignment scheme
affects mostly dwarf galaxies, overestimating their
circular velocities by a maximum of $\sim 20\%$. For larger galaxies
the effect decreases to less than $5\%$. This is due to the fact that
stripping only affects subhalos and they comprise only a minority
(about $20\%$) of the total halo population. In addition, only the
small number of subhalos which orbit close to their host halo's center
get significantly stripped and experience a substantial decline in their circular velocity.

Comparison of LV relations constructed using $\vacc$ one with the maximum  the
circular velocity (the dashed curve in Figure~\ref{fig:LV2}) and another with
velocities $\vacc$ estimated  at 10~kpc (the full curve)
indicates that this affects the largest halos the most. For example,
the $V_{10}$ velocity is almost a factor of two smaller than $V_{\rm
  max}$ for the group-size halos presented in the plot. Taking the
circular velocity at 10~kpc also makes the LV relation much less
curved as compared with the maximum velocity.  Below $\sim 80\kms$ the
maximum circular velocity of the DM halo happens near or within
10~kpc, which explains why the curve does not shift in this regime.

\subsubsection{The effects of cold baryons and adiabatic compression}
\label{sec:baryons}

\begin{figure}[htb!]
\epsscale{1.0}
\plotone{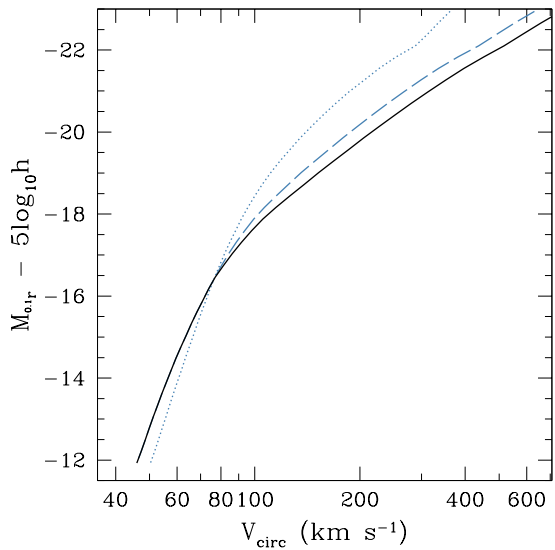}
\caption{Effects of baryons on the LV relation.  The dashed curve
  shows the circular velocity $\vten$ after adding the baryon mass at
  the center of each halo without any adiabatic contraction of the
  dark matter. The solid curve shows the result of implementing the
  correction due to the adiabatic contraction of the halos
  \citep{blumenthal86}. For reference, the dotted line shows the
  circular velocity of the DM measured at $10\kpc$ (assuming the
  baryons trace the DM distribution; same as solid
  curve in Figure~\ref{fig:LV2}). Baryons have little effect on
  dwarfs ($V_{\rm circ} < 100\kms$) since dwarfs are dominated by DM
  beyond a few kiloparsecs. Just adding the baryons in quadrature has
  the greatest effect with the adiabatic compression giving a 10-15\%
  correction for bright galaxies.}
\label{fig:LV3}
\end{figure}

Figure~\ref{fig:LV3} shows how cold baryons change the circular velocity at
a 10~kpc radius. Here we use two extreme approximations that bracket
the effect. The first approximation assumes that there is no change in
the distribution of the dark matter. All simulations so far indicate
that there is some compression. Hence, the no compression
approximation definitely underpredicts the circular velocity
$V_{10}$. The second approximation uses the adiabatic compression
model of
\citet{blumenthal86} which produces the largest increase in the
density of the dark matter. (The full and dashed curves were already
shown in Figure~\ref{fig:LV1}).
There are some differences between the LV
relations predicted by those approximations. However, the largest
effect is just adding the cold baryons in quadrature to the circular velocity
of the dark matter.  Adiabatic compression increases the circular
velocity even further, but the effect is relatively minor because the
fraction of cold baryons gets progressively smaller for larger
galaxies.  The amount of cold baryons used for the models is crucial for
this test. As we discuss in Section~\ref{sec:barfracBTF}, the
abundance matching predicts relatively small cold baryon masses for
dwarfs and giants, and this is why the adiabatic compression in
Figures~\ref{fig:Vcirc} and \ref{fig:LV3} is 10-20\% at the
most. Again, dwarfs below $\sim 80\kms$ are insensitive to the cold baryon
presence. Here the full curve (DM+baryons) is slightly below the
  $\vc$ of the dark matter curve because the latter includes a cosmological
fraction of baryons which trace the DM, most of which are assumed to
be blown out from galaxies (see Section~\ref{sec:procedure}).

\begin{figure}[htb!]
\plotone{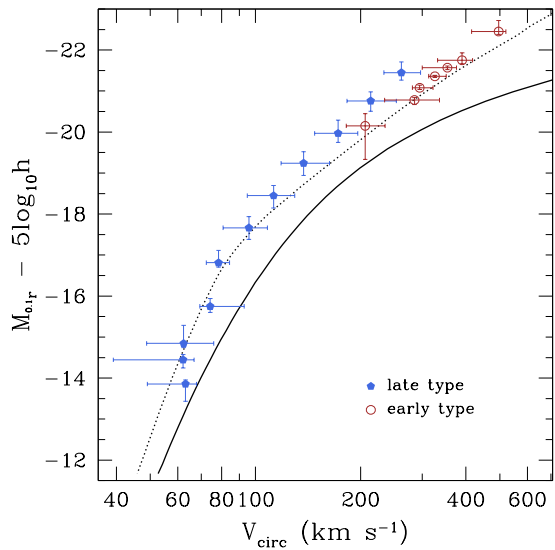}
\caption{Effect of excessive cold baryon mass. We  assume
  that {\it half} of the universal baryon fraction within each halo
  forms its galaxy.  Median values (solid curve) of $^{0.1}r$-band luminosity vs. circular
  velocity of our model galaxies measured at $10\kpc$ from the center
  and including the correction due to adiabatic halo contraction. For
  comparison, the
  dotted line and the symbols reproduce the model and the
  observational data shown in Figure~\ref{fig:LV1}. The model
  with $50\%$ cold baryon fraction systematically predicts galaxies
  that are too concentrated and fails to fit the observations.}
\label{fig:LV4}
\end{figure}

Figure~\ref{fig:LV4} shows the LV relation that we would obtain by assuming instead that half of all baryons within the virial radius or
equivalently 8\% of the virial mass are retained and are used to
build the central galaxy (while luminosity is not affected). Both the
shape and the normalization of the LV relation are incorrect, with the
circular velocities systematically larger than observations by up to $50\%$. Clearly,
it is difficult to obtain the observed LV relation assuming a constant
cold baryon fraction in the framework of the $\LCDM$ cosmology.

\subsubsection{The effects of including scatter in luminosity at fixed halo circular velocity}
\label{sec:scatter}

So far, the abundance matching procedure we have used assumed a monotonic one-to-one relation
between halo circular velocity and galaxy luminosity or stellar mass. This
assumption produces average relations that can be compared with the medians of the
observations. As shown by previous studies, a more detailed treatment
of the scatter between halo and galaxy properties may yield average
relations of the brightest galaxies that deviate significantly from the case with no
scatter. For instance, \citet{tasitsiomi04} showed that iteratively introducing
log-normal scatter of width $1.5$ mags in the assignment of luminosities to DM halos
produces an average TF relation with massive galaxies that are
brighter by about one magnitude compared to the monotonic
assignment. By treating the scatter analytically, \citet{behroozi10} found that performing
halo abundance matching using their preferred value of $0.16$ dex of log-normal scatter
reduces the average stellar mass assigned to massive halos with
total masses $> 10^{13}\Msun$ by up to $70\%$ (when binning using
virial mass) but does not affect less
massive galaxies below the knee of the stellar mass function.   

Appendix~\ref{sec:appendix} gives a detailed description of the method we employ to
introduce scatter in our model. In short, we obtain luminosities for each
of the galaxies in our sample by stochastically scattering the values
obtained in the monotonic assignment  while forcing the preservation
of the observed luminosity function. When scattering the values of
luminosity we do not constrain the shape of the probability
distribution (e.g., log-normal) or require its width to be constant
for all circular velocities. This is well justified since the shape of
the intrinsic scatter is more difficult to constrain observationally (e.g., due to observational systematics).

One parameter that our model does not currently predict is the width
of the probability distribution of luminosity at a fixed halo circular
velocity. This scatter originates from three main sources. The first is the
observational error in the determination of the true luminosities of
galaxies. Since we use the LF from the SDSS spectroscopic sample, these are the sum of the
errors in photometry plus the errors in the distances obtained from
spectroscopy. At the mean redshift of the sample these combined
errors are expected to be typically much less than $0.1$ mags. The
second source of scatter is the one present in the intrinsic relation between halo
circular velocity and galaxy luminosity due to variation in the physical processes of galaxy formation. \citet{Verheijen01} studied the nature of the
intrinsic scatter in the Tully-Fisher relation from HI observations
and obtained a value of $\sigma_{M_r}^{\rm int} = 0.38$ in the
{\it R}-band. This value is consistent with the distribution in the
observational samples used in this work. Lastly, since we assign
luminosities that are uncorrected for the inclination of disk
galaxies, we also need to include the scatter that results from the
distribution of dust extinction corrections observed in
SDSS. \citet{maller09} find a fit to this distribution as a weak function
of $r$-band luminosity and disk scale length. We adopt the value
$\sigma_{M_r}^{\rm ext} = 0.28$ they use for a galaxy with $M_K = -20$. We
neglect the errors due to photometry and distances and add the
remaining two contributions in quadrature to obtain $\sigma_{M_r}
\approx 0.5$, which we use to introduce scatter to the model galaxies
below the knee of the LF. Above this luminosity, where early-types
dominate, we assume that the lack of significant internal extinction
slightly reduces the scatter to $\sim 0.3$.

\begin{figure}[tb!]
\epsscale{1.0}
\plotone{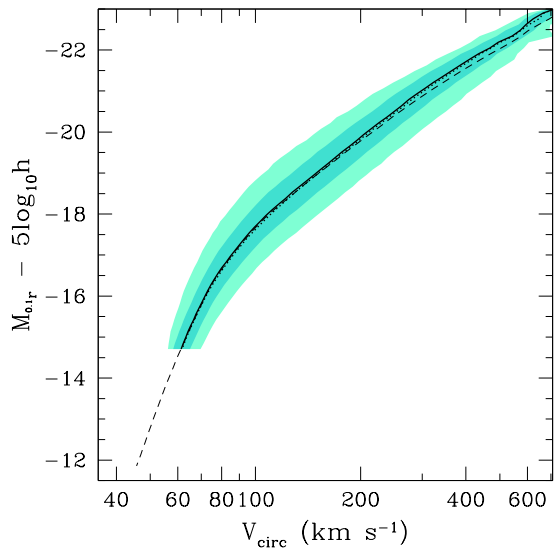}
\caption{The LV relation of the Bolshoi galaxies obtained using the
  stochastic assignment method described in Appendix~\ref{sec:appendix}
  to add scatter. The solid (dotted) lines show the median (average) of the circular
  velocity ($\vten$) in bins of $r$-band luminosity. The shaded areas
  encompass $68\%$ and $95\%$ of the galaxies in each luminosity
  bin. The dashed line shows the result of monotonic assignment with no
  scatter. The small, $\lesssim 0.3$ mag decrease of the average luminosity of the brightest
  galaxies is opposite in sign to the one obtained by binning in $\vten$. This is
 merely a result of binning bias due to the steepness of the velocity function.}
\label{fig:LV5}
\end{figure}

Figure~\ref{fig:LV5} shows the luminosity-binned distribution of model
galaxies in the LV relation obtained from the stochastic
HAM scheme and compares it to the monotonic assignment discussed in
Section~\ref{sec:TF}. Since we are left with a choice regarding which
quantity to average over, we choose to bin in $r$-band magnitude to be
consistent with the binning of the observations. The mean relation is almost identical to the
case with no scatter for galaxies below $200\kms$, while it becomes
brighter by up to $0.3$ magnitudes for more massive galaxies. Galaxies below $L^{\ast}$ show a
distribution of luminosities that is close to gaussian as far as 
2-$\sigma$ away from the mean but has a slightly longer bright tail. Galaxies brighter than $L^{\ast}$
show a trend of narrowing of the distribution as well as a skewness
that reduces the number of upscattered galaxies with increasing
luminosity. 

To check the consistency of our approach we also calculated the average LV relation of
our model galaxies obtained using the deconvolution method described
in \citet{behroozi10} and log-normal scatter. This procedure yields a deviation of the
mean relation for $\vc > 200 \kms$ towards higher luminosities that
depends on the assumed width of the scatter. Figure~\ref{fig:LV6} in
Appendix~\ref{sec:appendix} shows this effect for constant
$\sigma_{M_r} = 0.5$ (left panel) and $\sigma_{M_r}$ decreasing from
0.5 to 0.3 past $L^{\ast}$ (right panel). Even with a variable width
that mimics our approach, the luminosity-binned spread
obtained with the method of \citet{behroozi10} is unrealistically large at the
bright end. 

The differences between the results of the two methods actually reside
in the assumptions about the shape of the spread. Since our
stochastic assignment scheme does not constrain the scatter
distribution to be log-normal and centered on the monotonic relation,
the resulting skewness beyond $L^{\ast}$ allows it to preserve
the median LV relation of the scatterless sample. In addition, without
a skewed distribution it is extremely difficult to obtain a narrower
distribution of galaxies at the bright end of the LV relation.  

From this analysis we conclude that the introduction of scatter in luminosity at
a given halo circular velocity yields a median relation at the bright end
that is sensitive to the shape and width of the probability distribution
function used. The median LV relation is thus robust to uncertainties
in the nature of the scatter below the
shoulder of the LF, allowing for a direct comparison with observations. We prefer our
scatter method for two reasons. First, it {\it exactly} preserves the
luminosity function while the deconvolution method only does so
{\it approximately}. Second, it naturally produces an observed
luminosity-binned distribution that becomes narrower for the brightest
galaxies, in agreement with that expected from observations.

\subsection{Baryon fraction and the baryonic Tully-Fisher relation}
\label{sec:barfracBTF}

For the LV relation the cold baryons played an ancillary role: they
provided a correction to the circular velocity at 10~kpc. The
correction is small for galaxies below $100 \kms$. For large galaxies the
cold baryon contribution increases and typically is about half of the
mass within $10 \kpc$. Regardless of their role in the LV relation, baryons are one of
the prime subjects for the theory of  galaxy
formation. Unfortunately, accurate measurements of baryonic masses are also
prone to some uncertainties. Dynamical measurements of the baryonic
component are difficult because of dark matter-baryon
degeneracies \citep[e.g.,][]{Dehnen98}. In other words, the baryon mass 
depends on what is assumed about the dark matter. 
Population synthesis provides an independent estimate of the stellar
mass, but it has its share of complexities including the uncertainty
in the initial mass function. In addition to the stellar mass, most galaxies have an
important (if not dominant) fraction of their cold baryons in the form
of neutral hydrogen gas. For consistency, in this paper we make use of stellar
population synthesis estimates of stellar masses whenever possible.

The baryonic Tully-Fisher relation (BTF) is one way of displaying the amount of cold baryons in galaxies. The BTF relation has been investigated over the years
\citep{McGaugh00,Bell01,Verheijen01,McGaugh05,Stark09,McGaugh10}. Here we use
the recent observational samples of \citet{Stark09} and
\citet{Leroy08}, along with \citet{Verheijen01} and
the \citet{geha06} sample used for the LV relation.
\citet{Stark09} include gas-dominated spiral galaxies, which makes the
results much less sensitive to the uncertainties in the IMF. For
consistency, we calculate stellar masses for the \citet{Stark09} and \citet{geha06} samples using a
simple linear fit to the distribution of $V$-band mass-to-light ratios
vs. $(B-V)$ color shown in Figure 18 of \citet{Blanton07}: 
\begin{equation}
	M/L_V = 3.0(B-V) - 0.6 .  
\label{eq:MLratio}
\end{equation}
 Unlike the \citet{bell03} models, this relation fits well the measured
mass-to-light ratios of both blue and red galaxies in the SDSS. These estimates
are fully compatible with the stellar masses used in the stellar mass
function of \citet{li09} as part of our model.  \citet{Leroy08} present results based on the HI
Nearby Galaxy Survey (THINGS): \citet{Walter08}. The measurement of
luminosities in the infrared using {\it Spitzer} results in reliable
estimates of the stellar masses that are consistent with the
\citet{Blanton07} results. In this case we
adjust their stellar masses from the Kroupa to the \citet{chabrier03} IMF used
by \citet{Blanton07} by subtracting 0.05 dex. Selecting only galaxies
with high inclinations ($i>45 \degree$ or $b/a>0.7$) and better than 15\%
accuracy in the circular velocity data leaves a total of 161 galaxies. We also include the results of mass modeling of
the Milky Way and M31 \citep{Klypin02,Widrow05}. 

We also employ equation~(\ref{eq:MLratio}) to obtain stellar masses for the early-type galaxies. This ensures a fair comparison between the baryonic masses of  early- and late-type galaxies.

\begin{figure}[htb!]
\plotone{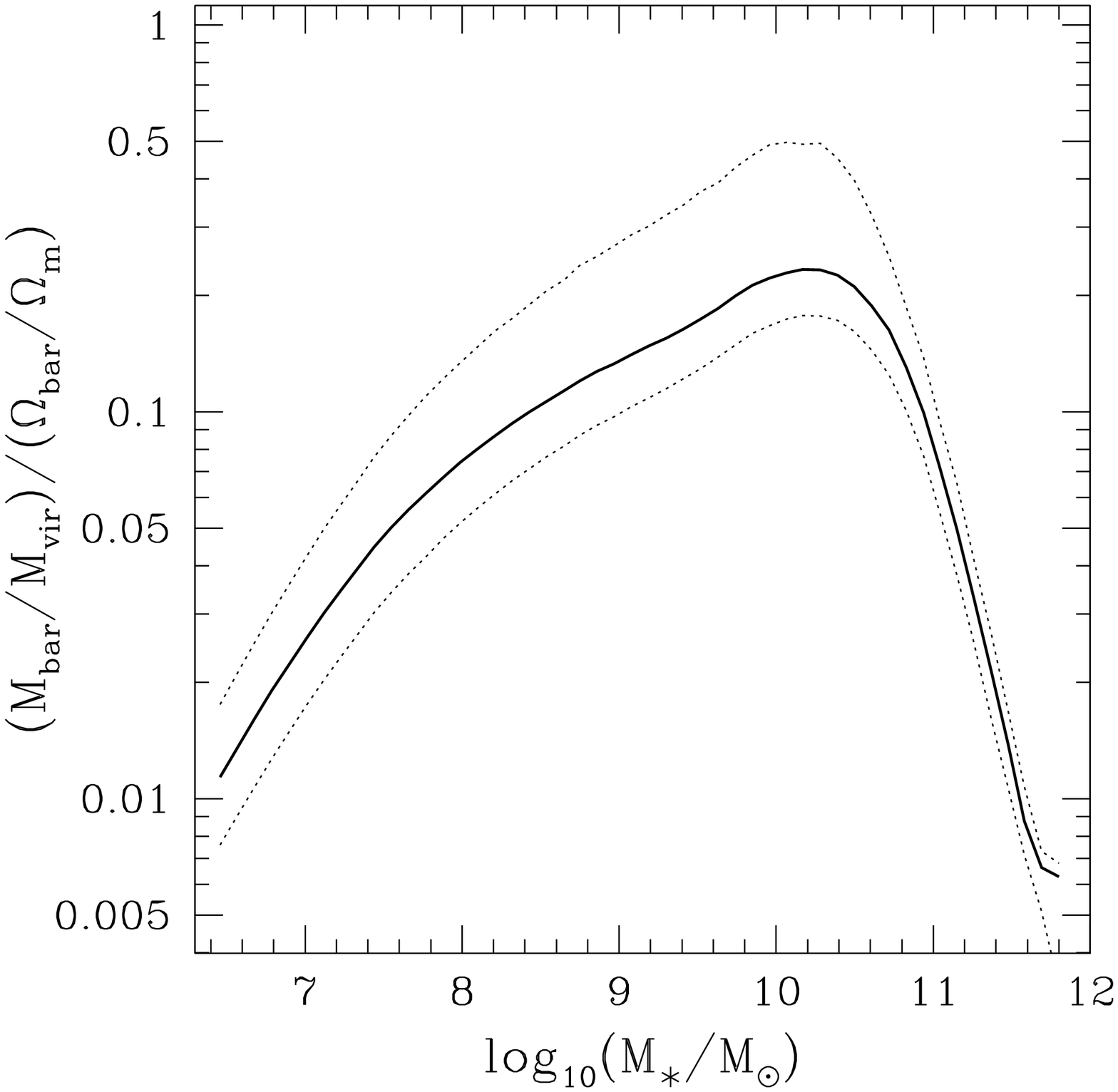}
\caption{Baryon fraction (in stars and cold gas) relative to the universal value as a function of stellar mass for the $\LCDM$ model using halo abundance matching. The solid and dashed lines show the median and 1-$\sigma$ scatter of the distribution respectively.}
\label{fig:barfrac}
\end{figure}

\begin{figure*}[htb!]
\epsscale{1.5}
\plotone{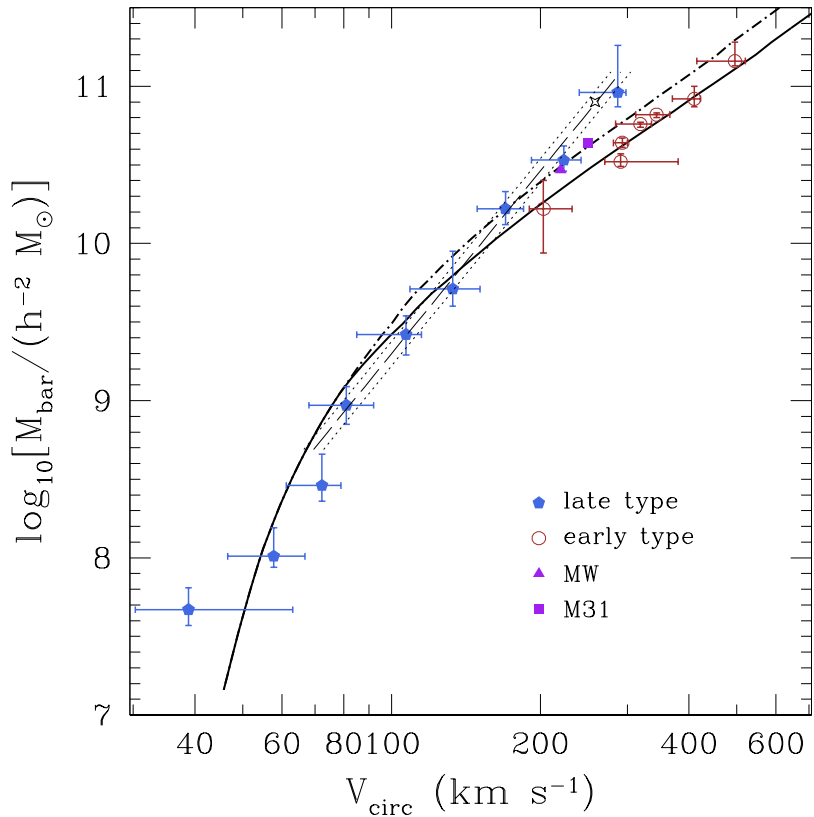}
\caption{Mass in cold baryons as a function of circular velocity. The
  solid curve shows the median values for the $\LCDM$ model using
   halo abundance matching and including adiabatic halo
   contraction. The cold baryonic mass
  includes stars and cold gas and the circular velocity is measured at
  a galactocentric distance of $10\kpc$. The dot-dashed curve shows the effect of neglecting halo contraction. For comparison we show the median and 1-$\sigma$
  scatter values of several binned galaxy samples.  Intermediate mass galaxies such as
  the Milky Way and M31 lie very close to our model results.}
\label{fig:BTF}
\end{figure*}

Figure~\ref{fig:barfrac} shows the cold baryon fraction relative to the
universal value as a function of stellar mass in our model. The cold baryon fraction
$f_b$ peaks at $\approx 0.2$ for the stellar masses typical of
Milky Way-type galaxies and sharply falls  on both sides of the
mass spectrum. Our results are broadly consistent with \citet{Guo10}.
We note that even the peak of $f_b\approx 0.2$ is almost a factor of two smaller 
than what a few years ago was considered a fiducial value
\citep{mo98}.

The baryonic Tully-Fisher relation is shown in Figure~\ref{fig:BTF}.
Theoretical estimates from abundance matching provide a good fit
to observational results for galaxies ranging from dwarfs with $\vcirc
\approx 60\kms$ to giants with $\vcirc \approx 500\kms$. In a
remarkable agreement with the LV relation result, the model with adiabatic
contraction seems to also provide a better fit to the BTF compared to
the model with no contraction. There is a hint
that observations show more baryonic mass for dwarfs below $\vcirc
=40\kms$ as compared to an extrapolation of the model. It is not clear whether this is a real problem because of
the uncertainties involved in the observations. FIrst, the small sample
size could produce biased results. Second, there is an uncertainty at the faint end
of the luminosity function. The results of abundance matching are
sensitive to the number density of galaxies with absolute magnitudes
$M_r > -14$, which is poorly constrained. 

As in the case of the LV relation, the model BTF relation agrees very well with the
average population of galaxies in each morphological regime. Below
$\simeq 200 \kms$ it follows late-type disks while it accurately
describes massive early types above this threshold. The observations
show no preference for a model with no halo contraction vs. one
with maximum contraction. Both cases fit well within the systematic and
statistical uncertainties in the observations.

Although S0 and elliptical galaxies seem to contain slightly less mass in cold
  baryons than massive spirals, there is also a hint that the bimodality observed in the observed LV
relation in Figures~\ref{fig:LV0} and \ref{fig:LV1} between early- and late-type galaxies
is not merely the result of a variation in the mass-to-light
ratio. Other authors have come to the same conclusions \citep[e.g.,][]{williams10,dutton10a,dutton10b}. This would imply that early types are not just the result of passive
fading of late-type disks but are fundamentally different. It also
requires that they inhabit deeper potential wells which
may be the result of different formation or environmental processes. These results have deep
implications for galaxy formation but in order to draw
conclusions we would need consistent stellar and gas mass estimates
for a larger sample of galaxies, which are not currently available.

\subsection{Galaxy Circular Velocity Function}
\label{sec:GVDF}

\begin{figure*}[htb!]
\epsscale{1.53}
\plotone{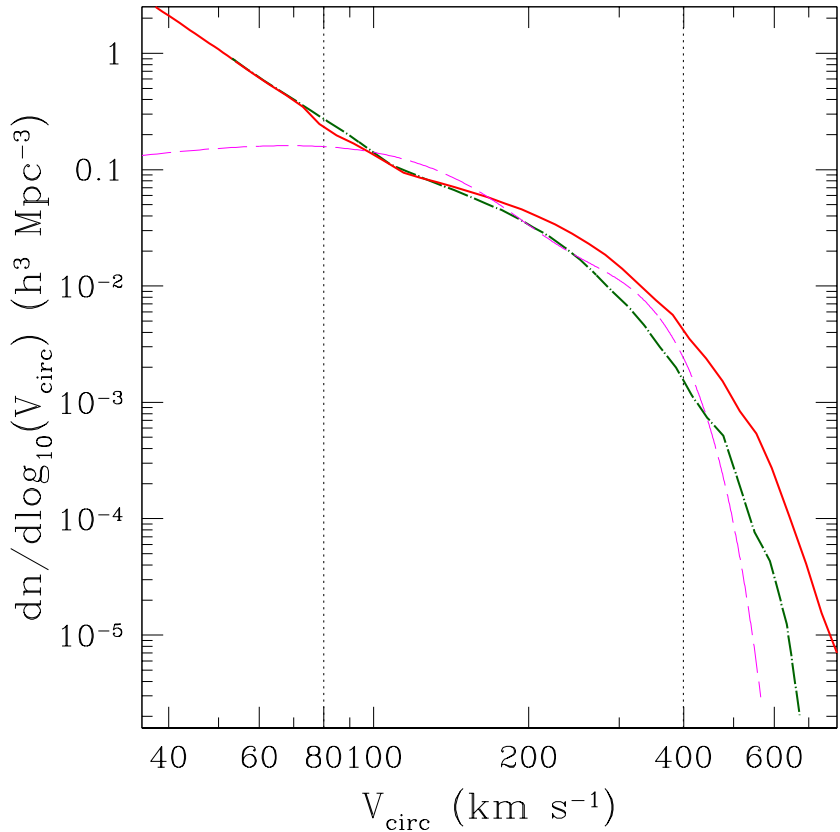}
\caption{Comparison of theoretical (dot-dashed and thick solid curves) and
  observational (dashed curve) circular velocity functions.  The
  dot-dashed line shows the effect of adding the cold baryons (stellar
  and cold gas components) to the central region of each DM halo and
  measuring the circular velocity at $10\kpc$. The thick solid line is the
  distribution obtained when the adiabatic contraction of the DM halos
  is considered. Because of uncertainties in the AC models, realistic
  theoretical predictions should lie between the dot-dashed and solid
  curves. Both the theory and observations are highly uncertain for
  rare galaxies with $\vcirc>400\kms$. Two vertical dotted lines divide the VF into three
  domains: $\vcirc>400\kms$ with large observational and theoretical
  uncertainties; $<80\kms<\vcirc<400\kms$ with a reasonable agreement,
  and $\vcirc<80\kms$, where the theory significantly overpredicts the
  abundance of dwarfs.}
\label{fig:GVF}
\end{figure*}

Projecting the distribution of galaxies in the LV plane onto the
luminosity axis produces the luminosity function, while projecting
onto the circular velocity axis yields the circular velocity function
(VF) of galaxies: the number-density of galaxies with given circular
velocity. From a theoretical cosmology point of view, the VF is an ideal
characterization because it does not include uncertain predictions for
the luminosity and requires relatively modest corrections for the baryonic masses. Unfortunately, it is more difficult to obtain it from
observations and so far, there have been only a few attempts to do so
\citep{Gonzalez00,Kochanek01,zavala09,chae10,zwaan10}.

For the theory the starting point is the velocity function of dark
matter halos \citep[e.g.,][]{klypin10}. For halos with
$\vcirc<500~\kms$ it is well approximated by a power-law $~n(>\vc)
\propto \vc^{-\alpha}$, where $\alpha \approx 3~$. This only applies
to velocities taken at the maximum of the circular velocity curves of
DM halos. For galaxies, the results must be adjusted to $V_{10}$ and
corrected for the dynamical effects of cold baryons.

The most recent measurement of the VF of nearby late-type galaxies was
obtained by \citet{zwaan10}. Their result is based on the blind HI
sample of the HIPASS survey, which is complete down to
$M_{\rm{HI}} = 5.5\times10^7 \Msun$ at a distance of $5\Mpc$
\citep{zwaan10}. Since gas-rich galaxies are thought to dominate at the low mass end, their
sample should provide an accurate measurement of the abundance of dwarfs
if these galaxies contain enough neutral gas to be detected.  To
obtain a galaxy velocity function for all morphological types we also
include the determination of the early-type VF done by \citet{chae10},
using the conversion between velocity dispersion and circular velocity
found in \citet{zwaan10}. Even though their VF was obtained indirectly
using the observed relation between luminosity and stellar velocity
dispersion, it agrees with previous direct measurements.

Figure~\ref{fig:GVF} shows the results, as well as the modified
Schechter fit to the VF of late-type galaxies \citep{zwaan10} and the
fit for early types found in \citet{chae10}.  
At intermediate to large masses ($80 < \vc < 400\kms$), where the
completeness of the surveys is hard to question, the VF of our model
sample reproduces the observed abundances reasonably well. The abundance
of MW-type galaxies is predicted to within $50\%$ when adiabatic
contraction is taken into account, and
within a few percent when no contraction takes place. From our earlier
analysis of the LV relation in Section~\ref{sec:results} we are led
to believe that halo contraction is needed to obtain the correct
position of elliptical and S0 galaxies in the plot. A more detailed
treatment of AC might be
necessary in order to better match the abundance of galaxies larger
than the Milky Way. Our model galaxy VF overestimates the abundance of
the most massive and rarest galaxies with $\vc > 400\kms$ regardless
of whether or not we implement the correction for contraction of the
halos. 
Most of these extremely bright galaxies inhabit the centers of 
clusters, where it is very likely that the simplistic observational 
estimate of $\vc$ is breaking down.
At small velocities ($\vcirc<80\kms$) the theory significantly
overpredicts the number of dwarfs. This ``missing dwarfs'' problem
remains unresolved in $\LCDM$ \citep{Tikhonov09a,zavala09,zwaan10}. It
should be noted that the variance of the velocity function of the
model galaxies below $60\kms$ in regions of radius $5\Mpc$ can be as
large as 1 order of magnitude. This shows that environmental bias may be an important
factor in explaining the underabundance of dwarfs in our model compared to HIPASS.   

\begin{figure}[tb!]
\epsscale{1.0}
\plotone{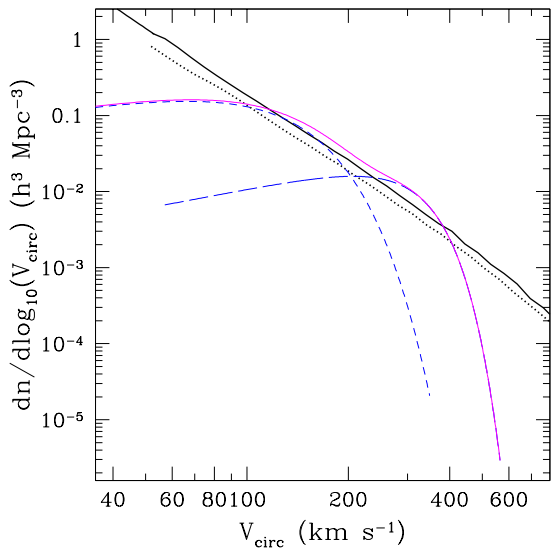}
\caption{Details of the velocity function.  The dotted line
  corresponds to the dark matter halo VF at $z=0$, while the thick
  solid line shows the distribution of galaxies obtained if the
  maximum rotation velocity of the halos is measured at its historical
  maximum (i.e. before accretion). Note that the total mass includes
  $17\%$ in baryons that behave like dark matter in dissipationless
  simulations. The short (long) dashed curve shows the Schechter fit
  for late (early) type galaxies. The thin full curve is the total
  observed VF.}
\label{fig:GVF1}
\end{figure}

To illustrate the effect that each of the steps in our procedure has
on the VF, we show in Figure~\ref{fig:GVF1} the VF of DM halos
only. It also shows that when the stripping due to the merger history
of each halo is considered, the halo VF does a slightly better job at
matching the abundance of galaxies.

As we previously noted, the corrections due to the presence of the
cold baryonic component affect dwarfs ($\vc < 100\kms$) very little,
resulting in a negligible shift in their abundance compared to that of
their host DM halos at the low-mass end of the VF in
Figure~\ref{fig:GVF}. One interpretation of this is that the dwarf
overabundance  problem cannot be resolved if both the LV relation and
the VF of dwarf galaxies are to be reproduced simultaneously. In other
words, these galaxies must undergo a process
that limits their abundance without changing their dynamical mass. The
first possible origin for the large discrepancy between our model
galaxies and the HIPASS VF could be observational bias. HIPASS is a
blind HI survey and does not detect gas-poor galaxies. Only if
gas-poor dwarf spheroidals dominate the galaxy population below $\sim
100\kms$ would it be possible to reconcile our results with the
survey. This is highly unlikely since this type of galaxies are only a
small fraction of the total dwarf population. On the other hand, if
the HIPASS HI mass detection limit ($5.5 \times 10^7\Msun$ at $5\Mpc$)
is relatively high at the distances where most of their sample is
found, incompleteness effects might explain the
discrepancy.

Assuming that the surveys are complete, a possible solution to the
problem is a mapping of {\it all} the dwarf galaxies below $50\kms$ to
DM halos in the range $50-100\kms$. This in turn implies that the
measured rotation curves of a large fraction of dwarfs must severely
underestimate the true maximum circular velocities of these
galaxies. The only possible explanation for this bias would be that
the optical and HI disk is truncated well inside the radius where the
rotation curve flattens
out. Another solution to the missing dwarf problem requires most
of these galaxies to have a low enough surface brightness in HI to be
undetectable in current surveys. This
would imply the existence of a large number of small halos containing
little or no neutral gas.

\subsection{Galaxy two-point correlation function}
\label{sec:CF}

The most important success of the halo abundance matching technique is
considered to be reproducing the observed galaxy clustering measured in the form of the
galaxy correlation function in its various forms, both in the local universe and at high redshift
\citep{tasitsiomi04,conroy06,Guo10,wetzel10}. Most of these works claimed
to match the observed clustering although they relied on
simulations with either very low resolution or outdated cosmological
parameters. The high resolution and large
volume of the Bolshoi simulation allow us to calculate the galaxy
two-point correlation function at a range of scales comparable to
the latest results from the final data release of the SDSS \citep{zehavi10}. In
addition, its up-to-date set of cosmological parameters allows for a
direct comparison between observations and the predictions of
$\LCDM$+HAM. The comparisons in this section do not make use of the
dynamical corrections that were necessary to obtain the LV relation and the
velocity function. Instead, the calculation of the galaxy correlation
function only requires the position and velocity information of the
halos in the simulation along with their luminosities obtained from
HAM. This makes the correlation function an even more robust prediction of the model.

In order to compare our model with observations, we use the most
recent measurement of the SDSS galaxy projected autocorrelation function done by
 \citet{zehavi10}. To make the best
comparison possible we use projected galaxy separations (a direct
observable)  and the same luminosity and projected radii
bins as \citet{zehavi10}. We also integrate along the line of sight
using the same distance bins while including the peculiar velocities
of the model galaxies in the redshift calculation. The integration
is traditionally performed to wash out the effects of redshift distortions. We limit the
calculation of the correlation function to distances below $30 \Mpch$
to avoid scales at which much of the power comes from long
waves that are absent in the simulation due to the finite box size. The small-scale correlation
function of DM halos is extremely sensitive to the abundance of
satellite halos near the centers of hosts. As a result of this, the
clustering in $N$-body simulations could be underestimated due to
artificial disruption of just a few satellites. Since it is beyond
the scope of this paper to perform a comprehensive study of this
effect, we choose to compare our model to galaxies in the range $-19 >
M_{r} - 5\log h > -22$. In this and all following sections we refer to
the $^{0.1}r$-band in shorthand as simply the $r$-band. 

\begin{figure*}[tb!]
\epsscale{2.0}
\plotone{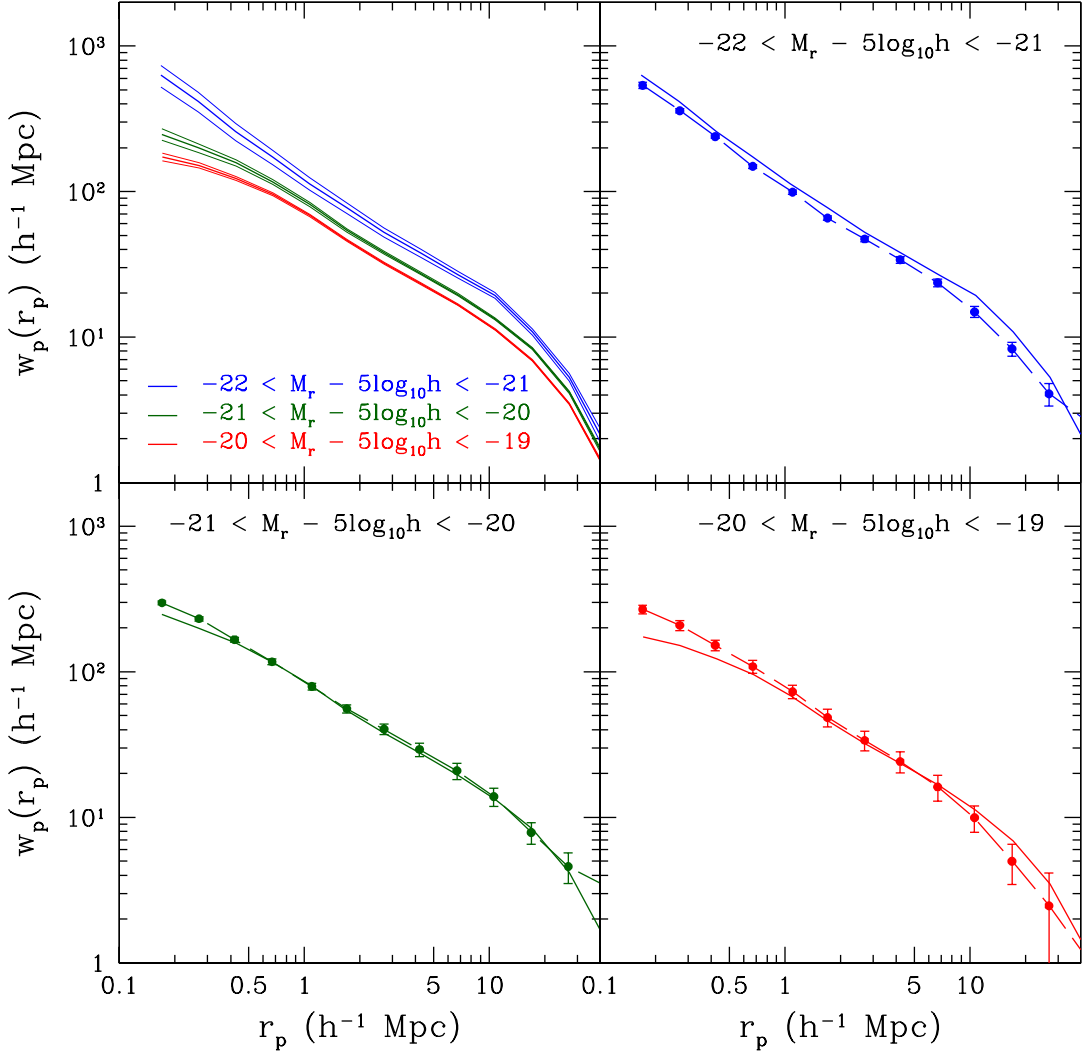}
\caption{The correlation function of the Bolshoi galaxies using HAM
  without scatter vs. the SDSS observations. {\it Top left:} correlation function of model galaxies in three magnitude bins showing the poisson uncertainties as thin
  lines. {\it Top right, Bottom left, Bottom right:} the clustering in
  each luminosity bin is compared to SDSS galaxies. Solid circles with error bars are the data
  from \citet{zehavi10}. $\LCDM$ + HAM does an excellent job at
  reproducing the shape and amplitude of the clustering of galaxies
  near the the knee of the luminosity function ($-20 > M_{r} - 5\log_{10} h
  > -21$). Brighter model galaxies are slightly more clustered than SDSS
  galaxies at large separations while faint ones underestimate the observed clustering
  at distances below $0.5 \Mpch$.}
\label{fig:CF}
\end{figure*}

Figure~\ref{fig:CF} shows the projected two-point correlation function of the
model galaxies in Bolshoi and compares it to the full SDSS sample
results. The clustering amplitude of the model galaxies is in excellent agreement with
observations for galaxies with luminosities around $L^{\ast}$ in the
range $-20 > M_{r} - 5\log_{10} h > -21$. Model galaxies with luminosities $-19 >
M_{r} - 5\log_{10} h > -21$ agree very well with the observations at scales
beyond $1-2 \Mpch$ where the clustering is dominated by halos of
different hosts (the so-called two-halo term). Below $1 \Mpch$ there is
a marked decline in the number of pairs as the separation decreases. For bright galaxies with $-21 >
M_{r} - 5\log_{10} h > -22$ the situation is different; $\LCDM$ + HAM
slightly overpredicts the clustering over all scales, with the disagreement increasing to
$\sim 30\%$ beyond $10\Mpch$. 

The discrepancy in the clustering of the faintest bin at small
separations may be a result of numerical effects such as artificial
disruption or halo misidentification in dense environments. A small
deviation at the closest separations ($r_p < 400 \kpch$) is likely to have the same
origin. Since the fraction of halos that are satellites decreases
sharply at large halo masses \citep[see][]{klypin10}, the model correlation
functions of the brightest galaxies do not suffer from these effects. Further
scrutiny is necessary to understand the origin of the effect and make
more robust comparisons with observations.

\subsubsection{Effect of scatter on the correlation function}

Since only the brightest galaxies in the LV relation are affected by scatter, the
obtained two-point correlation function of galaxies brighter than $M_r
\approx -22$ will be sensitive to the choice of scatter distribution. In the past
few years, some studies of the correlation function of DM halos have
suggested the scatter to be an essential ingredient in reproducing the
observations \citep[e.g.,][]{tasitsiomi04,wetzel10,behroozi10}.

\begin{figure*}[tb!]
\epsscale{2.0}
\plotone{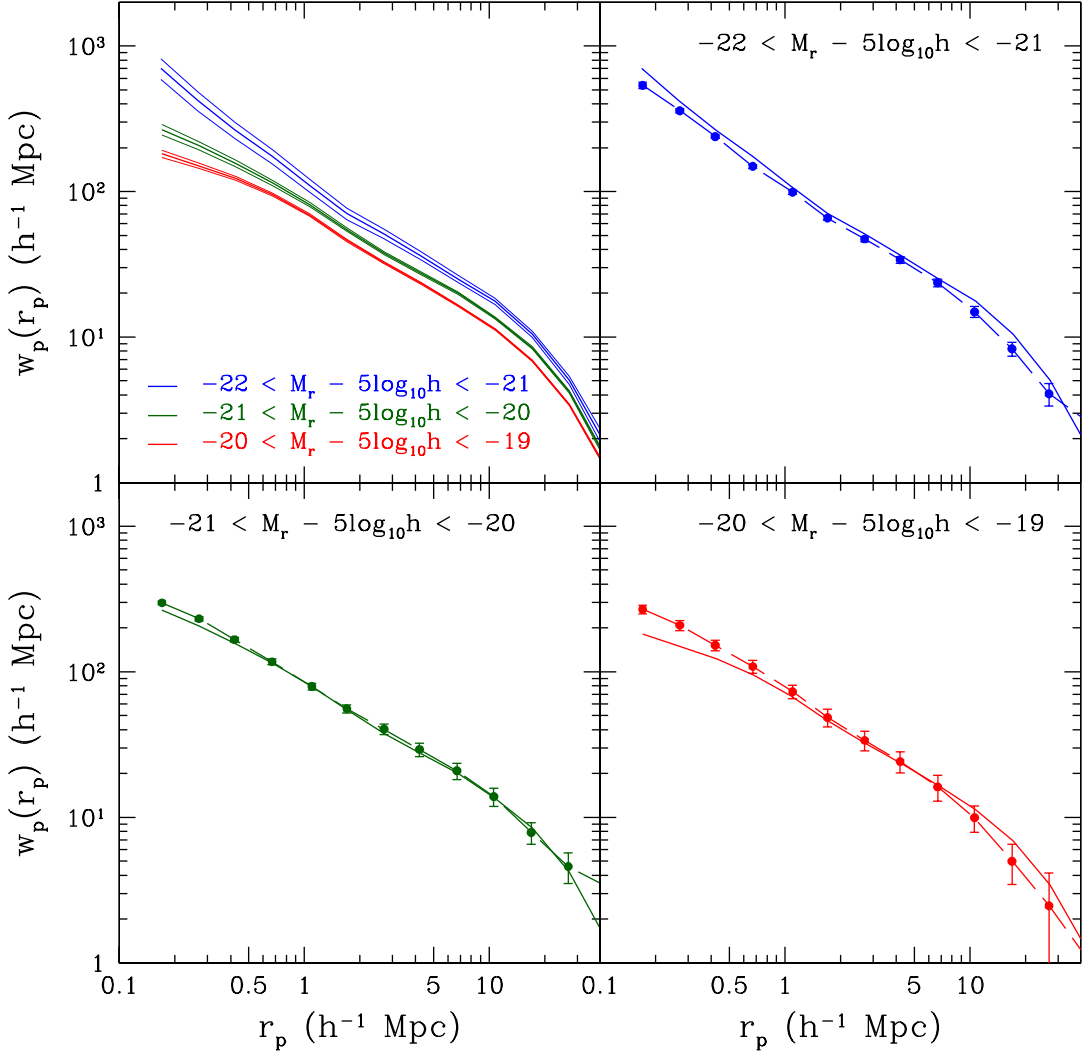}
\caption{Same as Figure~\ref{fig:CF} but including variable scatter
  in luminosity at fixed circular velocity using our stochastic
  abundance matching method. Galaxies with $-19 > M_r - 5 \log_{10} h > -21$
  are mostly unaffected while those in the brightest bin are slightly
  less clustered at all separations than in the case with no
  scatter. This results in a better agreement with the SDSS observations.}
\label{fig:CF3}
\end{figure*}

Figure~\ref{fig:CF3} shows the galaxy autocorrelation
function obtained for our model galaxies using the stochastic method
described in Appendix~\ref{sec:appendix} to perform HAM. Here we
assume the same distribution described in Section~\ref{sec:scatter}, with $\sigma_{M_r} \approx 0.5$
below $L^{\ast}$ and $\sigma_{M_r} \approx 0.3$ above. The correlation amplitude of
model galaxies fainter than $M_r - 5\log_{10} h = -21$ is essentially
unchanged compared to the monotonic assignment result shown in
Figure~\ref{fig:CF}. The clustering of the bright galaxies with $-22 <
M_r - 5\log_{10} h < -21$ shows a slight decrease at all separations except the smallest ones, and thus
better agreement with the SDSS data. The amplitude decreased due to the fact that the same galaxies now get
assigned to less massive halos on average, and these halos are less
clustered. This is consistent with the upward shift in the average luminosity of
the brightest galaxies in the LV relation
(Figure~\ref{fig:LV5}). The correlation functions
of galaxies in fainter bins are indistinguishable from those without scatter. 

In summary, applying our physically motivated scatter model maintains
(and even improves) the excellent agreement of the $\LCDM$+HAM model
with the observed galaxy clustering. The clustering of the most massive and rare
galaxies (those above $M_r \approx -22$) will be more sensitive to the addition of scatter and the model used to implement it. Once other
dominant sources of uncertainty in the simulations and observations are
better understood, a robust test of the cosmological model could be
done using these objects.

\section{Comparison with other results}
\label{sec:compare}

Our results are broadly consistent with \citet{Guo10}, who also use
the abundance matching technique. Specifically, in their Figure~6 they show
the stellar-mass -- circular velocity relation. The theoretical
velocities appear to be {\it smaller} than the observed circular
velocities for $\vcirc =100-150\kms$.  Although \citet{Guo10} did not
apply the necessary corrections discussed in our paper, they argue that
inclusion of the cold baryon mass may bring the theory into agreement with the
observations. As we show, indeed this is the case.

Incidentally, in the semi-analytic modeling paper \citep{guo11}
based on the Millennium-I and II simulations, the predicted 
angular correlation function of galaxies with $\log M_* < 10.77$ 
is significantly too high compared with SDSS data, especially 
at separations less than about 1 Mpc.  The authors
attribute this to the fact that the large $\sigma_8 = 0.90$ used in
the Millennium simulations produced too many massive halos 
that in turn host too many pairs of galaxies in their subhalos.

\citet{Dutton07} argue that the standard cosmological model with
adiabatic contraction and standard concentrations fails to
simultaneously reproduce
the observed LV relation and the luminosity function for late-type
galaxies. This conclusion is not compatible with our results.  A number of
assumptions made in \citet{Dutton07} are either outdated or need
corrections. For example, for their preferred model they use the
``standard concentrations'' of \citet{Bullock01}, which were based on
a simulation with $\sigma_8=0.9$, although they attempted to rescale
them to a cosmological model with the normalization
$\sigma_8=0.8$. 
The normalization of the current $\LCDM$ cosmological model is
$\sigma_8=0.82$ based on CMB and other data
\citep[e.g.][]{Jarosik10}, which results in halo concentrations that
are $\sim$30\% lower than what \citet{Dutton07} used in their
preferred model. In turn, this
reduces the dark matter circular velocities in the inner regions of halos
by about 15\%. Some of
the necessary corrections were discussed by \citet{Dutton07} and it was shown
that they substantially improve the fit of the TF relation. However,
the main difference is the treatment of the luminosity
function. \citet{Dutton07} use criteria obtained from SAMs to argue
that a model with halo expansion is necessary to match the LF. We
avoid such assumptions completely because our model reproduces the galaxy statistics automatically.

\citet{Gnedin07} studied structural properties of spiral galaxies and
compared them with theoretical predictions.  They also used
\citet{Bullock01} high concentrations as the ``standard'' model.  It
was concluded that the theory has problems and that adiabatic
contraction is the likely culprit. Another possible solution was to
lower the halo concentrations. Indeed, when \citet{Gnedin07} used
concentrations for a model with $\sigma_8=0.74$ as predicted by simple
theoretical arguments, they found that the theory gives an acceptable
fit to the data. The problem is that $\sigma_8=0.74$ is too low.
However, it seems that their analytical scaling with $\sigma_8$ was
not accurate enough: the concentrations {\it actually used} by
\citet{Gnedin07} are practically (within 3\%) the same as what we find
in $N$-body simulations for the Bolshoi $\LCDM$ model with
$\sigma_8=0.82$ \citep{klypin10}. In short, there seems to be no
contradiction between our results and \citet{Gnedin07} even when we
consider models with standard adiabatic contraction. More definite
conclusions require careful analysis and changes in the fraction of
cold baryons among other things.



\section{Discussion}
\label{sec:discussion}

In this paper we address one of the most difficult problems in
cosmology: is the standard cosmological $\LCDM$ model compatible with
observations when it comes to the prediction of the abundance and
properties of galaxies? Instead of focusing on traditional issues such
as the zero-point and the slope of the Tully-Fisher relation for
spiral galaxies, we work with a more generic luminosity-velocity (LV)
relation: a correlation of galaxy luminosity with the circular
velocity at a 10~kpc radius. We also investigate the (cold) baryonic mass -
velocity relation, which following tradition we call the baryonic
Tully-Fisher (BTF) relation, as well as the velocity function and the
two-point autocorrelation function of galaxies.
All these statistics encompass galaxies of different types - from dwarf
galaxies to normal spirals to giant ellipticals. These statistics --
in combination with the theoretical predictions of the cosmic
microwave background and the abundance and
properties of dark matter halos -- are major tests for the validity of the $\LCDM$ model.

We use the abundance matching technique to assign luminosities to
halos predicted by cosmological simulations. We also use abundance
matching to assign stellar and cold baryon masses.  We find that all
three statistics -- the LV and BTF relations, and the velocity function -- provide
reasonably good fits to observations for galaxies ranging over 10
magnitudes in luminosity and for circular velocities from $80\kms$ to
$400\kms$. By construction, our models fit the observed luminosity and
stellar mass functions. Since they are based on the Bolshoi simulation
\citep{klypin10}, they also fit known properties of dark matter halos
including the halo mass function and the dependence of halo
concentration on mass. In addition, in this paper we show that halo abundance
matching also yields the correct clustering
properties of bright galaxies. In short, {\it we have a model, that fits --
  at least on average -- all the basic statistics of galaxies with
  $\vcirc > 80\kms$ considered at a $\sim$10~kpc scale}.

Matching theory with observations requires a  careful consideration of
many different effects and application of different corrections. These
effects were considered both for observations and for the theory.  On the
observational side, we compiled a representative sample of galaxies
with measured circular velocities. Velocities were either asymptotic
values (``flat part'' of rotation curves) for spirals or measurements
at $\sim 10$~kpc radius for S0s and Es.  We do not use fits (such as
power-laws) to the data but instead work directly with the distributions. We do not apply
morphological corrections of the TF relation (e.g., differences
between Sb and Sa galaxies) because those corrupt the bright end of
the LV relation. Since the Tully-Fisher luminosities are corrected to face-on,
we de-correct the magnitudes of galaxies for the effect of 
internal absorption to make them consistent with the measurement of
the luminosity function. 

For the theoretical predictions we try to make all the possible corrections to
mimic the observational situation.  For example, we do not use virial
masses of halos because virial radii are too large compared with
the typical distances at which rotational velocities of observed galaxies
are measured. We do not assume a particular shape for the halo density profiles:
they are measured directly in the simulations.  The simulations required for this type of
analysis should have a very high resolution so that subhalos are also
resolved. This allows us to avoid using intermediate steps such as
the Halo Occupation Distribution or the Conditional Luminosity
Function, which are often applied to low-resolution simulations. The
Bolshoi cosmological simulation \citep{klypin10} provides high quality
results resolving distinct halos and subhalos down to the completeness
limit of $\vcirc =50\kms$.

The observations should be taken
cautiously since each has a different
degree of accuracy. The LV relation is the most accurate because it is easier to
measure luminosities than to estimate stellar masses, which require
additional modeling and assumptions. This is why we consider the LV
relation as our prime target.  The velocity function is the least
reliable since observations are still at the very early stages. The
completeness of the HIPASS VF is very uncertain. Just the fact that
the detection limit is quoted at $5\Mpc$ shows that the accuracy of
the HI mass function is not very high. This is why we treat the
results on the velocity function for $\vcirc > 80\kms$ as a ``pass'' for
the theory in spite of some deviations such as at $130\kms$. More
accurate treatment of these gas-rich galaxies may also change the situation: after all,
changes in abundances and velocities by $\sim 10\%$ may (or may not)
resolve the discrepancies. 

It is more difficult to reconcile the theory and observations at
smaller velocities. Indeed, at $\vcirc =50\kms$ the formal
disagreement is almost a factor of ten. This is the only serious problem 
that we find when matching galaxies with dark matter
halos. A similar problem on somewhat smaller scales was reported by
\citet{Tikhonov09a}, who studied the population of dwarfs in the $\sim
10$ Mpc region centered on the Milky Way galaxy. \citet{Tikhonov09b}
argue that Warm Dark matter may be the solution to the problem.

We introduce a simple scatter model that is well motivated and
  preserves the agreement with the LV relation and the correlation
  function of galaxies in the SDSS. The introduction of scatter has some complications. Observed deviations from the median relations
seem to have a systematic component: early-type galaxies are
systematically below the median LV relation and gas-rich spirals are
above it. It seems likely that the LV relation -- like the
color-magnitude diagram -- has a bimodal structure. In this case, no
simple gaussian spread can explain the whole diagram. However uncertain,
the spread must be explained.  One approach might be to match halos
separately to red and blue galaxies, for example using local density
as well as luminosity.  Ultimately it will be necessary to find the
real source of the dynamical bimodality and to measure it
observationally.

Although our $\LCDM$+HAM prescription makes simplifying assumptions regarding the
distribution of baryons in DM halos, it yields results that are
compatible with more detailed dynamical models. Our model
predicts $\vc$ values that differ by less than 5\% from
\citet{dutton10b} for massive disk galaxies without halo
contraction. Given that \citet{dutton10b} include a large set of
observational constraints on the radial distribution of baryons, the consistency with our results is evidence of the robustness of our model\footnote{\citet{dutton10b} used LV relations to find that late- and early-type galaxies are best fitted by halo expansion and halo contraction (\`a la \citealt{gnedin04}), respectively. As mentioned above, we do see a systematic difference between early- and late-types in our assembled data sets that might be explained by some combination of differences in halo contraction and halo masses (Figures~\ref{fig:LV1} and \ref{fig:BTF}). However, given the observational uncertainties and the non-differentiation between early and late-types in our models, we cannot yet provide a precise interpretation of the bimodality.}.

Abundance matching is a very successful way to make
predictions about how on average galaxies can inhabit dark matter
halos. It gives up solving the most difficult and the most important
problem: how galaxies form inside dark matter halos. It simply assumes
that the stellar mass and luminosity monotonically (or possibly with
some scatter) scale with the
circular velocity.  Bluntly speaking, it assumes that the maximum circular
velocity of a halo determines the properties of the galaxy hosted by
that halo. Remarkably, this can reproduce some basic
environmental relations such as the morphology-density relation and the
dependence of galaxy clustering on the luminosity of galaxies
\citep[e.g.,][]{conroy06} because of the correlation of environment
with the average halo mass \citep[][]{Sheth01,Sheth04}. However, there
are potential issues with abundance matching. It is not clear how it can
explain dependencies on environment even if galaxies are selected with
the same $r$-band luminosity or the same stellar mass
\citep{Hogg04,Wel08}. Modeling of the
bimodality in the LV relation (the apparent differences between early
and late type galaxies) is another problem to address. It will be interesting
to see how much better the results will be from more sophisticated
abundance matching including galaxy color and local density -- as, e.g., in \citet{tasitsiomi04} -- and from semi-analytic
modeling based on the Bolshoi simulation.  This work is in progress.

Disk formation and
  semi-analytic models still struggle to simultaneously reproduce the TF relation (a
  subset of the LV relation) and
  the abundance of galaxies \citep[e.g.,][]{benson03,monaco07,benson10}. In
  this paper we have shown that our model is successful at this task. Simultaneously reproducing the luminosity function and the LV
  relation depends critically on implementing each of the steps in
  Section~\ref{sec:procedure} to obtain the properties of the galaxies that inhabit
  $\LCDM$ halos. For
  example, Figure~\ref{fig:LV4} shows how assuming an incorrect value
  for the baryon
  fraction (as in \citet{mo98}) can lead to an LV relation that is in striking disagreement with
  the observations. In the more recent semi-analytic model of
  \citet{benson10} the circular velocities of galaxies are $40-50
  \kms$ larger than observed at any luminosity. This may be explained
  by the fact that their baryon fraction is about $20\%$ larger than
  ours for
  Milky Way-mass galaxies and about an order of magnitude larger than
  our result for the most
  massive ellipticals as well as dwarfs. Our model shows that galaxies with masses larger
  that the Milky Way have circular velocities that are extremely
  sensitive to the baryon content within their
  optical radius which may explain why SAMs overpredict the circular velocity.

Previous works based on abundance
  matching were successful at reproducing the statistics of the integrated
  properties of galaxies (such as clustering as a function of
  luminosity and redshift) but made no attempt to include their internal
  and baryonic structure \citep[e.g.,][]{Kravtsov04,conroy06,wetzel10,behroozi10,Guo10}. In this paper we show that making robust dynamical
corrections to the structure of halos obtained in simulations gives
the correct galaxy scaling relations. These corrections include adding the cold baryonic component and measuring
$\vcirc$ at $10\kpc$. Previous studies using HAM
could not include these corrections partly because
simulations lacked the large dynamic range necessary to form the largest
halos and resolve substructure adequately. The large box size and very
high resolution of the Bolshoi simulation makes it possible to obtain
good statistics of even the largest clusters and resolve the structure of dwarf halos.

\section{Conclusions}
\label{sec:conclusions}

Here is a short summary of our results:

\begin{itemize}

\item In combination with previous results, we conclude that the
standard $\LCDM$ model in conjunction with halo abundance matching can simultaneously fit reasonably well the
main global statistics of galaxies: the luminosity function, the
stellar mass function, the Luminosity-Velocity relation, the baryonic
Tully-Fisher relation, the abundance of galaxies with circular
velocities $\vcirc > 80\kms$, as well as the clustering properties of 
bright galaxies as a function of luminosity.

\item There are systematic deviations in the LV relation with S0 and
elliptical galaxies located about 1 magnitude below late types in the
LV relation. Massive early types contain less baryonic mass than late
types at the same circular velocity. 

\item The range of the effect of contraction of the DM halos due to
  baryon infall brackets the
  observations. The LV relation shows preference for a model with moderate
  contraction, as predicted by \citet{gnedin04}. This result is compatible
  with the observed velocity function of galaxies. 

\item There seems to be an overabundance of model galaxies by a factor of $\sim 10$
  compared to observed dwarf galaxies with
$\vcirc < 50\kms$. This is a serious problem for the $\LCDM$ model: galaxies
with these circular velocities cannot be affected much by ``normal''
physical processes (e.g., supernovae feedback or reionization of the
Universe) proposed for the solution of the satellite
problem at $\vcirc \lesssim 30\kms$. However, the observational
results on the abundance of dwarf galaxies still need to be improved.

\item Including scatter in luminosity at fixed $\vc$ using our physically
  motivated scatter model maintains the agreement of the model LV
  relation with observations. 

\item The correlation function of the
  model Bolshoi galaxies matches very well the observations of bright
  galaxies. The agreement improves when implementing our scatter prescription for all but the brightest galaxies, where a better understanding of the uncertainties is necessary to make a fair test of $\LCDM$ using halo abundance matching. 

\end{itemize}

\acknowledgements We thank A. Dutton, T. Davis, F. Prada, R. Wechsler,
A.~Kravtsov, M. Williams, Fill Humphrey and P. Behroozi for helpful conversations,
M. Williams for providing his data in advance of publication, and
Kyoko Matsushita for providing data in electronic form. We also thank
the anonymous referee for many insightful comments regarding the
manuscript. We acknowledge support of NSF grants at NMSU and NASA 
and NSF grants at UCSC.  Our simulations and analysis were done using NASA
Advanced Supercomputing (NAS) resources at NASA Ames Research Center.
AJR was supported by National Science Foundation Grants AST-0808099
and AST-0909237.


\appendix

\title{APPENDIX}

\section{Halo abundance matching including stochastic scatter}
\label{sec:appendix}

To perform halo abundance matching while including scatter in
luminosity as a function of circular velocity we perform the following procedure. For
brevity, wherever we use $M_r$ we refer to $M_{^{0.1}r} - 5\log_{10} h$.

\begin{enumerate}

\item Start with the monotonic assignment described in
  Section~\ref{sec:procedure}: Order the list of DM halos from largest to smallest
  $\vc$. Using the integral LF, solve for the unique luminosities of the
  galaxies that have the same number densities as the DM halos in
  Bolshoi. This matching gives the monotonic relation $M_r^{\rm mono}(\vc)$. 

\item For the halo with the largest $\vc$, draw a luminosity value $M_r$ at random from a Gaussian distribution of width $\sigma$ centered at a point
  1-$\sigma$ brighter than the value of $M_r^{\rm mono}$ for that halo. Mathematically, this is equivalent to $M_r =
  M_r^{\rm mono}(\vc) - \sigma + \mathcal{G}(0,\sigma)$, where
  $\mathcal{G}(0,\sigma)$ is a random realization of a Gaussian
  probability distribution with standard deviation $\sigma$ centered at zero. 

\item If the randomly drawn $M_r$ is brighter than the brightest
  galaxy in the LF, another draw is performed and the process is
  repeated until a suitable value is found.

\item The random $M_r$ is compared to the list of available $M_r$
  values in the list. The closest available value becomes the luminosity of that halo. If
  the value is already taken, another random draw is performed until
  an available one is found. This luminosity value is flagged to prevent
  it from being used again for another halo. This step ensures that
  the observed luminosity function is preserved by only assigning each
  luminosity once.

\item In order to avoid having unassigned values of $M_r$ in the list on the bright
  tail of the distribution, we check whether there is
  any unassigned luminosity which is more than 3-$\sigma$ brighter
  than $M_r^{\rm mono}$. If such unassigned value exists, the
  next halo in the list is assigned to it. Since we are stepping along the list of luminosities as we assign them, the process is intrinsically asymmetric and there
  will always be some leftover $M_r$ values that get assigned in this
  step. The offset used in step 1 ensures that these make up only a
  few percent of the sample. 

\item Repeat steps 2-5 for each halo in the ordered list until the
3-$\sigma$ faint tail of the Gaussian for a given halo reaches the completeness
limit of the sample as defined in Section~\ref{sec:sim}. The procedure
is stopped at this point to prevent placing a hard constraint on the
faint end of the LF where it is most uncertain. The faintest halos (up
to 0.5 magnitudes brighter than the cutoff) are removed from the sample to insure the
preservation of the LF throughout. 

\end{enumerate}

In spite of the fact that a constant Gaussian distribution is used in
the algorithm, the final distribution of $M_r$ for a given circular
velocity $\vc$ is not a Gaussian and the width of the $M_r-\vc$
relation is not constant. This happens because of the asymmetry in the
distribution of the galaxies: there are always more galaxies with
smaller luminosities than with larger ones. Thus, the distribution and
the spread of the final $M_r-\vc$ relation are the result of a convolution
of a Gaussian distribution with the luminosity function. 

The effect of the asymmetry of the LF
is more pronounced for the brightest galaxies in the sample and
accounts for the small shift in the median compared to the monotonic
assignment. The algorithm also gives a natural narrowing of the
distribution of luminosities as $\vc$ increases. This occurs because
in the exponential tail of the LF the number of available $M_r$ values
changes rapidly across the width of the Gaussian. Increasingly fewer
available values on the bright side of the median force the selection
of most values to take place in a narrower interval on the faint side. 

Since our assignment method reduces the width of the
obtained distribution of luminosities, we choose the input value
$\sigma = 0.7$. This yields a distribution with $\sigma_{M_r}
\approx 0.5$ below $\sim 200 \kms$ and gradually decreasing to $\sigma_{M_r} \approx 0.3$ above $\sim 300 \kms$.

\begin{figure}[htb!]
\epsscale{0.75}
\plotone{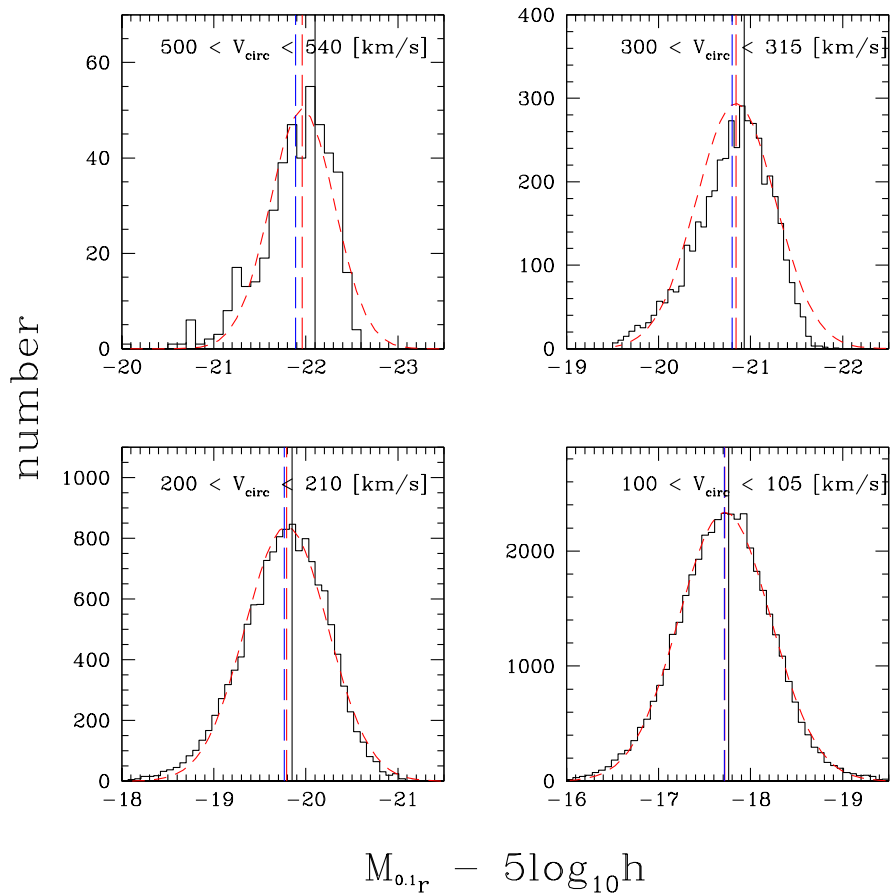}
\caption{The distribution of model galaxies obtained using the
  stochastic abundance matching method. Each panel shows one of four
  representative circular velocity ($\vten$) bins. The vertical dashed lines in each
  panel show the median and average while
  the vertical solid line shows the average value that was assigned in the monotonic
  scheme. The dotted lines show gaussian fits to each distribution. As galaxies become brighter (from
  bottom right to top left), the distribution narrows and becomes
  slightly skewed towards the faint tail.}
\label{fig:scatter1}
\end{figure}

Figure~\ref{fig:scatter1} shows the distribution of luminosities in
four bins of circular velocity, $\vten$. Below $200 \kms$, galaxies
show a near-normal distribution that is centered very close to the
values obtained from the monotonic assignment without scatter. For
$\vcirc > 250 \kms$ the distributions get progressively more skewed as
galaxies move from the bright to the faint tail.  The spread of the distribution of
luminosities increases with $\vc$: the widths are $\sigma_{M_r} \approx
0.50$, $0.45$, $0.43$ and $0.35$ for the bins centered at $102.5$,
$205.0$, $307.5$ and $520.0 \kms$ respectively.

\begin{figure*}[hb!]
\epsscale{1.05}
\plottwo{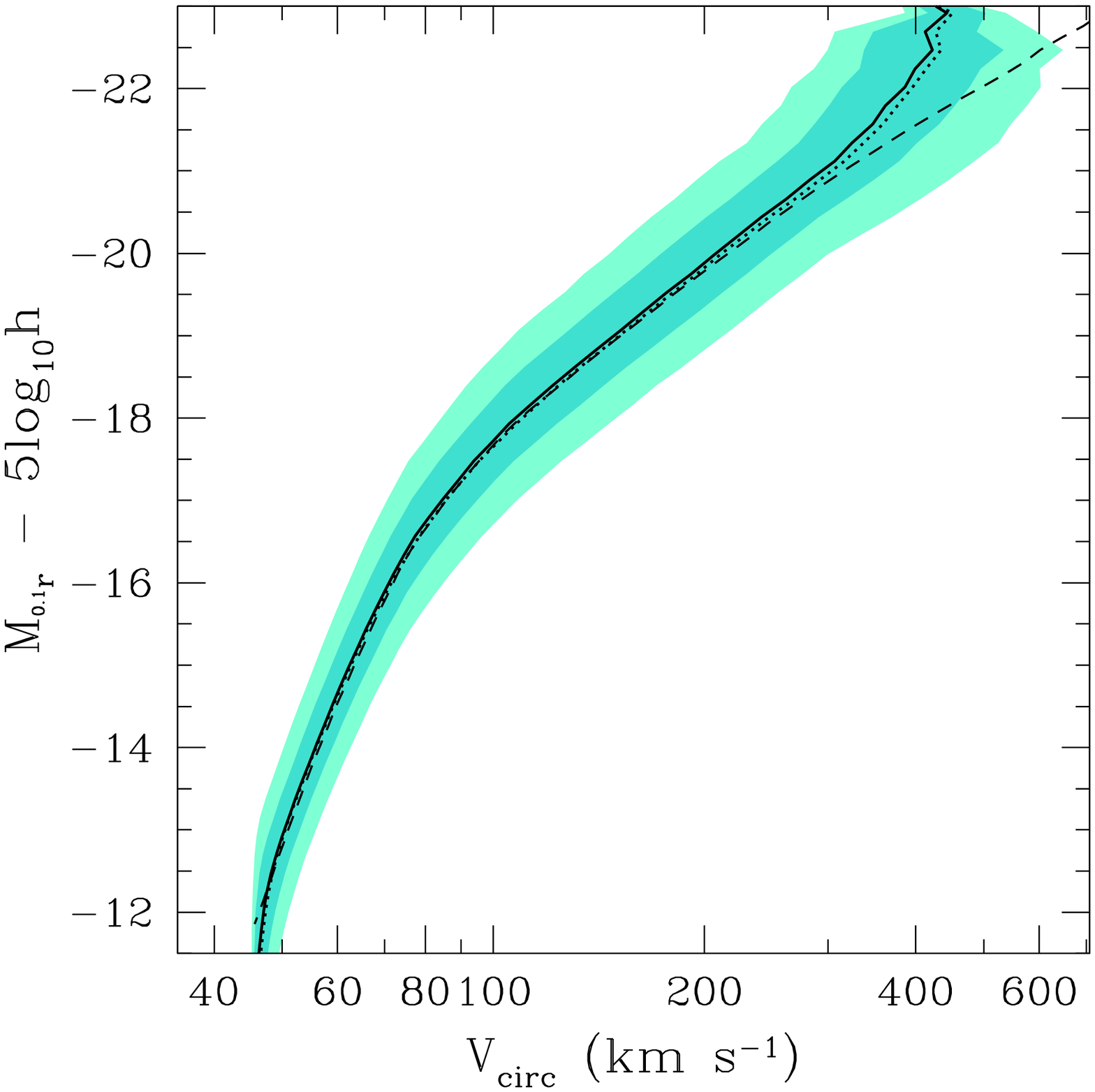}{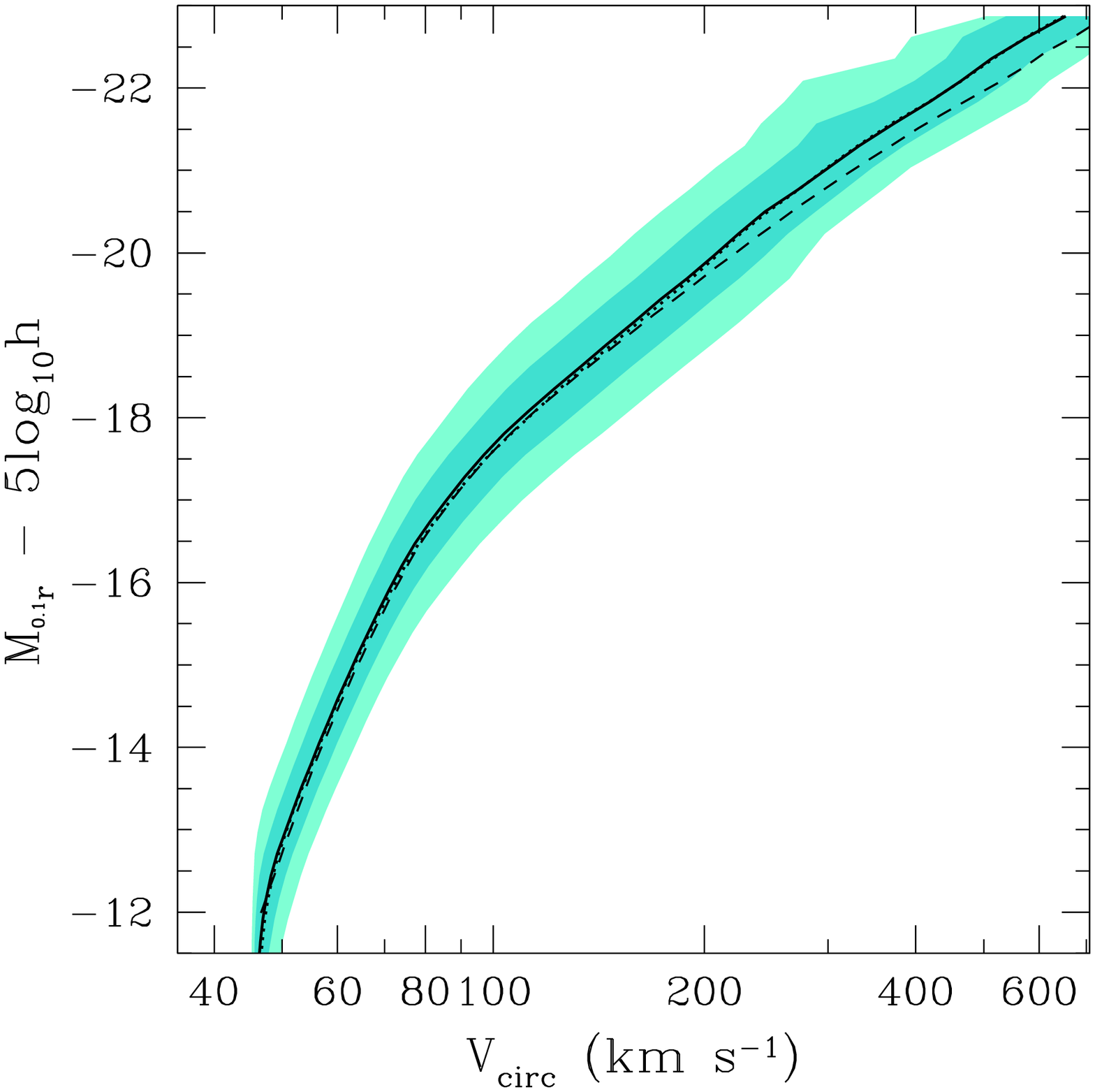}
\caption{The LV relation of the Bolshoi galaxies obtained using the
  deconvolution method assuming log-normal scatter. The solid (dotted)
  line shows the median (average) of the circular velocity in bins of
  $r$-band luminosity. The shaded areas encompass $68\%$ and $95\%$ of
  the galaxies in each bin. The dashed line shows the result of
  monotonic assignment with no scatter. {\it Left:} result of using a constant scatter width $\sigma_{M_r} = 0.5$. The median relation deviates by up to ~1
  magnitude for the brightest galaxies compared with the monotonic
  result shown as a dashed line. {\it Right:} result of using a width
  $\sigma_{M_r} = 0.5$ below $L^{\ast}$ and  $\sigma_{M_r} = 0.3$
  above. Although the median only deviates by a small amount, the spread in luminosity of the distribution of galaxies with $M_r - 5\log_{10} h < -20$ is considerably larger than that obtained using stochastic HAM.}  
\label{fig:LV6}
\end{figure*}

Figure~\ref{fig:LV6} shows the LV relation obtained with the scatter
model of \citet{behroozi10} in the case of constant scatter (left
panel) as well as assuming the same variable width used in our model:
$\sigma_{M_r} \approx 0.5$ for $\vc < 250 \kms$ declining to $\sim
0.3$ for larger $\vc$. Evidently, the median of the distribution in
luminosity stays relatively unchanged only when forcing a small scatter
width at the bright end. Even when we impose the same variable width used in the
stochastic HAM scheme, the resulting spread is too large for the brightest galaxies.

\section{Early-type data}
\label{sec:ETGtable}

Table~\ref{table:ETGdata} gives some of the properties of our early-type sample along with the source of data for each galaxy. 

\begin{deluxetable}{lccccccccccccccc}
\tablewidth{0pt}
\tabletypesize{\footnotesize}
\tablecaption{Luminosity and circular velocity data for nearby
  early-type galaxies.}
\label{table:ETGdata}
\tablehead{Name & Type & $M_B$ & $\pm$ & $\log_{10}\left(\frac{M_*}{M_\odot}\right)$ & $V_{10}$ & $+$ & $-$ & $V_{15}$ & $+$ & $-$ & $V_{20}$ & $+$ & $-$ & probe & ref.}
\startdata
NGC 4889 &      E    & $-22.61$ &       0.16 & 11.87 &  520 &   60 &    50 &    --- &   --- &   ---  &   --- &  --- &    --- & stars &   T+07   \\
NGC 4874 &      E    & $-22.51$ &       0.16 & 11.65 &  500 &   30 &    50 &    510 &   60 &    70 &     --- &  --- &    --- & stars &   T+07   \\
NGC 0315 &      E    & $-22.27$ &       0.25 & 11.66 &  520 &   10 &    5  &    530 &   30 &    10 &     550 &  35  &     20 & stars &   K+00   \\
NGC 1316 &	S0   & $-22.20$ &       0.19 & 11.46 &	381 &	29 &	29 & 	466 &	52 &	52 &	361  &	49  &	49   & X-ray & 	NM09 	\\
NGC 4839 &      E    & $-22.14$ &       0.15 & 11.52 &  385 &   65 &    30 &    400 &   80 &    50 &     425 &  105 &     70 & stars &   T+07   \\
NGC 0057 &      E    & $-22.06$ &       0.19 & 11.56 &  491 &   20 &    49 &    494 &   19 &    52 &     495 &  22  &     52 & X-ray &   O+07   \\
NGC 4555 &      E    & $-22.05$ &       0.21 & 11.50 &  614 &   58 &    62 &    606 &   61 &    66 &     598 &  63  &     74 & X-ray &   OP04   \\
NGC 4952 &      E    & $-21.75$ &       0.19 & 11.37 &  405 &   10 &    15 &    435 &   20 &    30 &     455 &  25  &     45 & stars &   T+07   \\
NGC 6407 &      E/S0 & $-21.73$ &       0.26 & 11.51 &  445 &   25 &    35 &    450 &   50 &    40 &     460 &  60  &     60 & stars &   MB01   \\
NGC 4472 &      E    & $-21.64$ &       0.12 & 11.41 &  415 &   40 &    40 &    --- &   --- &   --- &    --- &  --- &     --- & stars &   MB01  \\
NGC 7626 &      E    & $-21.56$ &       0.20 & 11.41 &  420 &   9  &    10 &    410 &   15 &    15 &     390 &  20  &     15 & stars &   K+00   \\
NGC 5044 &	E    & $-21.49$ &	0.27 & 11.31 &	273 &	17 &	17 &	328 &	11 & 	11 &	365  &	13  &	  13 & X-ray &	NM09 \\
NGC 4486 &      E    & $-21.40$ &       0.16 & 11.31 &  503 &   47 &    34 &    --- &  --- &   --- &     --- &  --- &    --- & GCs   &   M+11   \\
NGC 3923 &	E    & $-21.35$ &	0.43 & 11.26 &	365 &	28 &	28 &	324 &	40 & 	40 &	324  &	40  &	  40 & X-ray &	NM09 \\
NGC 4816 &      E/S0 & $-21.35$ &       0.23 & 11.19 &  300 &   45 &    40 &    300 &   50 &    40 &     310 &  50  &     40 & stars &   T+07   \\
NGC 1395 &      E    & $-21.32$ &       0.17 & 11.26 &  374 &   17 &    17 &    --- &   --- &   --- &    --- &  --- &     --- & X-ray &   NM09  \\
NGC 4944 &      S0   & $-21.31$ &       0.32 & 11.19 &  275 &   2  &     2 &    280 &    2 &    2  &     280 &  2   &     2  & stars &   T+07   \\
NGC 4382 &      S0/a & $-21.31$ &       0.16 & 11.12 &  260 &   19 &    19 &    260 &   19 &    19 &     198 &   51 &      51 & X-ray &   NM09  \\
NGC 4649 &      E    & $-21.29$ &       0.16 & 11.27 &  425 &   10 &    10 &    436 &   21 &    22 &     444 &  35  &     37 & X-ray &   HB10   \\
NGC 4374 &      E    & $-21.25$ &       0.12 & 11.26 &  382 &   20 &    20 &    386 &   20 &    20 &     393 &  20  &     20 & PNe &   N+11   \\
NGC 4827 &      E/S0 & $-21.25$ &       0.23 & 11.20 &  350 &   50 &    30 &    --- &   --- &   --- &    --- &  --- &     --- & stars &   T+07  \\
NGC 0128 &      S0   & $-21.20$ &       0.20 & 11.18 &  370 &   14 &    14 &    361 &    15 &    15 &    --- &  --- &     --- & stars &   W+09  \\
IC 1459  &      E    & $-21.17$ &       0.21 & 11.26 &  338 &   53 &    53 &    258 &    53 &    53 &    282 &   45 &     45 & X-ray &   NM09  \\
NGC 4957 &      E    & $-21.14$ &       0.19 & 11.21 &  325 &   2  &     2 &    310 &   10 &    2  &     295 &  15  &     2  & stars &   T+07   \\
NGC 4261 &      E    & $-21.04$ &       0.20 & 11.20 &  362 &   27 &    29 &    338 &   27 &    29 &     332 &  30  &     33 & X-ray &   HB10   \\
NGC 6703 &      E/S0 & $-21.07 $ &      0.20 & 11.11 &  220 &   20 &    15 &    --- &   --- &   --- &    --- &  --- &     --- & stars &   K+00  \\
NGC 3665 &      S0   & $-21.06$ &       0.23 & 11.13 &  423 &   52 &    52 &    404 &    32 &    32 &    404 &   32 &      32 & X-ray &   NM09  \\
NGC 5846 &      E    & $-21.04$ &       0.24 & 11.19 &  340 &   5  &     5 &    --- &   --- &   --- &    --- &  --- &     --- & stars &   K+00  \\
NGC 4908 &      E    & $-21.00$ &       0.15 & 11.19 &  320 &   40 &    40 &    --- &   --- &   --- &    --- &  --- &     --- & stars &   T+07  \\
NGC 0720 &      E    & $-20.99$ &       0.18 & 11.18 &  317 &   12 &    13 &    323 &   13 &    13 &     330 &  11  &     12 & X-ray &   HB10   \\
NGC 7796 &      E    & $-20.99$ &       0.20 & 11.15 &  308 &   46 &    26 &    305 &   37 &    28 &     297 &  35  &     26 & X-ray &   O+07   \\
NGC 4365 &      E    & $-20.98$ &       0.18 & 11.15 &  333 &   26 &    26 &    333 &   26 &    26 &     --- & ---  &   ---  & X-ray &   NM09   \\
NGC 3607 &      E/S0 & $-20.91$ &       0.20 & 11.07 &  278 &   28 &    28 &    278 &   28 &    28 &    265 &   28 &      28 & X-ray &   NM09  \\
NGC 1399 &      E    & $-20.88$ &       0.19 & 11.11 &  430 &   25 &    30 &    --- &   --- &   --- &    --- &  --- &     --- & stars &   K+00  \\
NGC 3585 &      E    & $-20.77$ &       0.22 & 10.99 &  295 &   46 &    46 &    --- &   --- &   --- &    --- &  --- &     --- & X-ray &   NM09  \\
NGC 2974 &      E    & $-20.76$ &       0.20 & 11.06 &  304 &   10 &    10 &    --- &   --- &   --- &    --- &  --- &     --- & gas   &   W+08  \\
NGC 5084 &      S0   & $-20.73$ &       0.21 & 11.05 &  282 &    8 &     8 &    --- &   --- &   --- &    --- &  --- &     --- & stars &   W+09  \\
NGC 6771 &      S0/a & $-20.72$ &       0.17 & 10.92 &  340 &   18 &    18 &    331 &    18 &    18 &    320 &   23 &      23 & stars &   W+09  \\
NGC 4807 &      E/S0 & $-20.64$ &       0.23 & 10.99 &  295 &   30 &    15 &    --- &   --- &   --- &    --- &  --- &     --- & stars &   T+07  \\
NGC 4931 &      S0   & $-20.62$ &       0.23 & 11.02 &  280 &    5 &    15 &    275 &    10 &    25 &     270 &  15  &     30 & stars &   T+07   \\
NGC 0821 &      E    & $-20.58$ &       0.21 & 10.88 &  182 &   13 &    13 &    --- &   --- &   --- &    --- &  --- &     --- & stars &   FG10  \\
NGC 1332 &      E/S0 & $-20.56$ &       0.22 & 10.94 &  291 &   9  &    10 &    291 &    9  &    10 &     291 &  9   &     10 & X-ray &   HB10   \\
IC 0843 &       S0   & $-20.55$ &       0.19 & 10.83 &  380 &   10 &    5  &    340 &    20 &    10 &     320 &  25  &     15 & stars &   T+07   \\
ESO151-G004 &   S0   & $-20.48$ &       0.26 & 11.10 &  291 &   19 &    19 &    308 &    19 &    19 &    295 &   25 &      25 & stars &   W+09  \\
NGC 4494 &      E    & $-20.40$ &       0.17 & 10.77 &  198 &   10 &    10 &    188 &   14 &    14 &     184 &  18  &     18 & PNe   &   N+09   \\
NGC 4869 &      E    & $-20.38$ &       0.15 & 10.91 &  280 &   50 &    30 &    --- &   --- &   --- &    --- &  --- &     --- & stars &   T+07  \\
NGC 4636 &      E    & $-20.38$ &       0.17 & 10.85 &  430 &   16 &    16 &   491  &   22  &    22 &    538 &   29 &      29 & X-ray &   NM09  \\
NGC 1032 &      S0/a & $-20.33$ &       0.21 & 10.82 &  270 &   20 &    20 &    --- &   --- &   --- &    --- &  --- &     --- & stars &   W+09  \\
NGC 4697 &      E    & $-20.20$ &       0.18 & 10.73 &  235 &   15 &    10 &    234 &   17 &    19 &     231 &  23  &     26 & PNe   &   dL+08  \\
IC 4045 &       E    & $-20.19$ &       0.15 & 10.85 &  390 &   40 &    30 &    --- &   --- &   --- &    --- &  --- &     --- & stars &   T+07  \\
\\
\\
\\
\\
NGC 3203 &      S0/a & $-19.89$ &       0.25 & 10.47 &  229 &    7 &     7 &    --- &   --- &   --- &    --- &  --- &     --- & stars &   W+09  \\
NGC 3379 &      E    & $-19.84$ &       0.11 & 10.69 &  206 &   31 &    13 &    192 &   42 &    19 &     181 &  48  &     22 & PNe   &   dL+09  \\
NGC 3957 &      S0/a & $-19.24$ &       0.20 & 10.38 &  199 &   13 &    13 &    --- &   --- &   --- &    --- &  --- &     --- & stars &   W+09  \\
NGC 4710 &      S0/a & $-19.10$ &       0.21 & 10.12 &  182 &   10 &    10 &    --- &   --- &   --- &    --- &  --- &     --- & stars &   W+09  \\
NGC 4469 &      S0/a & $-18.77$ &       0.17 & 10.22 &  182 &   13 &    13 &    --- &   --- &   --- &    --- &  --- &     --- & stars &   W+09  \\
\enddata
\tablecomments{The galaxies are sorted by absolute magnitude $M_B$,
which is corrected for Galactic extinction;
the quoted errors include the statistical uncertainties in distance and photometry.
The stellar masses are based on $B-V$ colors, and correspond to a Chabrier IMF (see text).
Circular velocities $V_{10}$, $V_{15}$, and $V_{20}$ are measured at 10, 15, and 20~kpc,
respectively, and are in units of \kms.
References:
dL+08: \citet{deLorenzi08};
dL+09: \citet{deLorenzi09};
FG10: \citet{Forestell10};
HB10: \citet{Humphrey10};
K+00: \citet{Kronawitter00};
M+11: \citet{Murphy11};
MB01: \citet{Magorrian01};
N+09: \citet{Napolitano09};
N+11: \citet{Napolitano11};
NM09: \citet{Nagino09};
OP04: \citet{OSullivan04};
O+07: \citet{OSullivan07};
T+07: \citet{Thomas07};
W+08: \citet{Weijmans08};
W+09: \citet{williams09a}.
}
\end{deluxetable}

\end{document}